\documentclass[11pt]{article}
\usepackage[dvipsnames]{xcolor}
\usepackage{amsmath,amsfonts,amssymb,amsthm,amssymb,bbm,bm}
\usepackage{pdfsync}
\usepackage{cite}
\usepackage{graphicx,import}
\usepackage[latin1]{inputenc}
\usepackage{hyperref}
\usepackage{empheq}
\usepackage[all]{xy}
\usepackage{stmaryrd}
\usepackage{rotating}
\usepackage{color}  
\usepackage{slashed}
\usepackage{cancel}
\usepackage{array, makecell} %
\usepackage{comment}
\usepackage{tikz-cd}
\usepackage{hhline}
\usepackage{comment}
\usepackage{mathrsfs}
\usepackage{enumitem}
\usepackage{stackengine}
\stackMath

\newcommand{\p}{\partial}

\newcommand{\bit}{\begin{itemize}}
\newcommand{\eit}{\end{itemize}}
\newcommand{\bd}{\begin{description}}
\newcommand{\ed}{\end{description}}

\newcommand{\bc}{\begin{center}}
\newcommand{\ec}{\end{center}}

\newcommand{\be}{\begin{equation}}
\newcommand{\ee}{\end{equation}}
\newcommand{\bea}{\begin{eqnarray}}
\newcommand{\eea}{\end{eqnarray}}
\newcommand{\bs}{\begin{subequations}}
\newcommand{\es}{\end{subequations}}

\def\p{\partial}

\def\bz{\bar{z} }

\newcommand{\questionequal}{\mathrel{\stackon[1pt]{=}{?}}}

\def\dcft{\Delta}
\newcommand{\Ocft}[1]{O_{#1}}
\newcommand{\fOcft}[1]{\widetilde{O}_{#1}}
\newcommand{\cftm}[2]{O_{#1}^{(#2)}}
\newcommand{\fcftm}[2]{\widetilde{O}_{#1}^{(#2)}}

\def\dccft{\delta}
\newcommand{\Occft}[1]{\mathcal{O}_{#1}}

\newcommand{\Ocarr}[1]{\mathscr{O}_{#1}}
\newcommand{\Ocarrft}[1]{\widetilde{\mathscr{O}}_{#1}}
\newcommand{\carrm}[2]{\mathscr{O}_{#1}^{(#2)}}
\newcommand{\fcarrm}[2]{\widetilde{\mathscr{O}}_{#1}^{(#2)}}

\begin{document}

\begin{titlepage}

\unitlength = 1mm
\ \\
\vskip 2cm
\begin{center}

{\LARGE \bf{Infinite towers of 2d symmetry algebras\\
\vspace{12pt} from Carrollian limit of 3d CFT}}
\date{}

\vspace{0.8cm}
Leonardo Pipolo de Gioia$^1$ and Ana-Maria Raclariu$^2$
\vspace{0.7cm}\\
{\small{${}^{1}$\textit{ICTP South American Institute for Fundamental Research
IFT-UNESP, \\
Sao Paulo, SP Brazil 01440-070}, \\
\vspace{5pt}
${}^2$\textit{King's College London, Strand, London WC2R 2LS, United Kingdom}} }
\vspace{20pt}

\begin{abstract}
\vspace{10pt}
We consider the Carrollian limit of OPE blocks of scalar primaries, spin-1 currents and the stress tensor in 3-dimensional conformal field theory (CFT$_3$). We demonstrate that these OPE blocks decompose into OPE blocks of towers of $\mathfrak{sl}(2,\mathbb{C})$ modes labeled by negative integer dimensions. We present two applications of this construction. We first show that the current-scalar-scalar and stress tensor-scalar-scalar OPE blocks in CFT$_3$ reduce to $\mathfrak{sl}(2,\mathbb{C})$ OPE blocks from which towers of conformally soft photon and graviton theorems can be derived. We then show that the CFT$_3$ OPE blocks of, respectively, current and stress tensor components dual to positive helicity gluons and gravitons in AdS$_4$ become $\mathfrak{sl}(2,\mathbb{C})$ blocks of conformally soft gluons and gravitons which imply the $S$ and $w_{1 + \infty}$ algebras of celestial CFT.

\end{abstract}
\vspace{0.5cm}
\end{center}
\vspace{0.8cm}

\end{titlepage}
\tableofcontents

\newpage

\section{Introduction}

It is now well known that the infrared sector of gauge theories and gravity in $(3+1)$-dimensional (4d) asymptotically Minkowski spacetime exhibits features of 2-dimensional conformal field theory (CFT$_2$). The observation that the 4d Lorentz group acts as the global conformal group on 2d cuts of null infinity $(\mathscr{I}^{\pm})$ led to early explorations of the decomposition of unitary Poincar\'e algebra representations into $\mathfrak{sl}(2,\mathbb{C})$ ones \cite{Dirac:1936fq, deBoer:2003vf}. Only more recently were these ideas applied in a scattering context. Upon decomposing the plane waves into Lorentz boost eigenstates (also known as conformal primary wavefunctions) one can reexpress scattering amplitudes of massless (or massive) particles in terms of observables transforming like correlation functions of primary operators in a CFT$_2$ \cite{Pasterski:2016qvg,Pasterski:2017kqt}.  In this new basis, the leading and subleading soft graviton theorems  can be recast as Ward identities of a Kac-Moody current and the stress tensor in a CFT$_2$ \cite{He:2014laa, Kapec:2014opa}, thereby promoting $\mathfrak{sl}(2,\mathbb{C})$ to an infinite dimensional Lie algebra. 

Quite remarkably, further symmetries can be found by analyzing the collinear limit of gravitons in a conformal primary basis: certain primary descendants \cite{Pasterski:2021fjn,Pasterski:2021dqe} of conformal primary gravitons with negative integer scaling dimension were shown to generate a $w_{1 + \infty}$ algebra \cite{Guevara:2021abz, Strominger:2021lvk}. From a 4d perspective, the $w_{1 + \infty}$ generators are related to modes in an all-order soft expansion of the graviton \cite{Hamada:2018vrw,Jiang:2021ovh}. Furthermore, an action of this algebra on the gravitational phase space can be constructed by analyzing the asymptotic expansion of the Weyl tensor to all orders in $r^{-1}$ \cite{Freidel:2021higher}. A similar infinite higher-spin 2d symmetry algebra, now known as the $S$ algebra, appears in Yang-Mills theory \cite{Guevara:2021abz, Strominger:2021lvk, Freidel:2023gue,Banerjee:2023bni}. In self-dual theories, the origin of these symmetry algebras has been attributed to integrability, which becomes manifest in twistor space \cite{Adamo:2021lrv,Kmec:2024nmu,Kmec:2025ftx}. The fate and implications of these 2d higher-spin symmetry algebras in more general theories are currently largely unknown. 

The goal of this paper is to show that all these symmetries appear when carefully examining the Carrollian limit of operator product expansions involving a conserved current and the stress tensor in 3-dimensional Lorentzian conformal field theories (CFT$_3$). The only assumptions involved in our derivation are that the CFT$_3$ contains a conserved spin-1 and/or spin-2 current and that the associated three-point coefficients are related to the scalar primary three-point function dual to a contact diagram in AdS$_4$. In this case, the towers of symmetries appear in a certain kinematic limit of Witten diagrams of gluons and gravitons in AdS$_4$. 

Introducing spacetime coordinates $x = (t,z,\bar{z})$ in the CFT$_3$, the Carrollian limit is defined by setting $t = c u$ and taking $c \rightarrow 0$ keeping $(u,z, \bz)$ fixed \cite{Levy-Leblond:1965dsc}. In this limit, the $\mathfrak{so}(3,2)$ algebra is enhanced to a conformal Carroll algebra $\mathfrak{ccarr}(3)$ \cite{Duval:2014uva,deGioia:2023cbd} which is isomorphic to the $\mathfrak{bms}_4$ algebra. This limit can be studied in different contexts with independent applications (for a recent review, see \cite{Bagchi:2025vri}). Here, we are interested in its application to flat-space holography. In particular, the isomorphism between $\mathfrak{ccarr}(3)$ and $\mathfrak{bms}_4$ suggests that the AdS/CFT correspondence should admit a well-defined flat space limit. The flat space limit has a long history \cite{Susskind:1998vk,Giddings:1999jq}, with most practical applications employing the Mellin representation of CFT correlators from which flat space scattering amplitudes (typically of massive particles) can be obtained in a certain scaling limit \cite{Penedones:2010ue}. In this framework, the leading soft theorems were first shown to arise from Ward identities associated with vector currents in the CFT \cite{Fitzpatrick:2012cg}. The emergence of soft theorems from a flat-space limit of AdS/CFT was further elaborated on in \cite{Hijano:2020szl}, while more recently, the emergence of flat-space scattering amplitudes/Carrollian correlators/celestial amplitudes from 3d CFT correlators in a Carrollian (or bulk point) limit of CFTs was established in \cite{PipolodeGioia:2022exe, de2023celestial, Bagchi:2023fbj, Bagchi:2023cen, deGioia:2024yne, Alday:2024yyj, Lipstein:2025jfj, Kulkarni:2025qcx}.

Building on observations of \cite{Kapec:2017gsg}, we showed in \cite{de2023celestial} that both the 4d leading and subleading soft gravitons can be identified with respectively the sub-leading and leading coefficients in an expansion of the CFT$_3$ \textit{shadow} stress tensor in powers of $u$, \textit{after} taking the Carrollian limit. It is then natural to expect that the whole tower of soft graviton modes, and hence the $w_{1+\infty}$ algebra generators, is encoded into coefficients in a Laurent expansion of the CFT$_3$ stress tensor (or shadow stress tensor) components $T_{zz} \equiv T^+, ~ T_{\bz\bz} \equiv T^-$ ($\widetilde{T}_{zz} \equiv \widetilde{T}^+, ~ \widetilde{T}_{\bz\bz} \equiv \widetilde{T}^-$), schematically\footnote{This identity should be understood as holding when inserted into correlation functions.}
\begin{equation}
\label{eq:stress-tensor-exp}
   \lim_{c \rightarrow 0}  T^+(cu, z, \bz)  \sim   \mathscr{T}^+(u,z,\bz)  = \sum_{n} u^n  \mathcal{G}^+_{3+n}(z,\bz) .
\end{equation}
By construction, $\mathcal{G}^+_{3 + n}(z,\bz)$ transform as $\mathfrak{sl}(2,\mathbb{C})$ primaries of dimension $\delta = 3 + n$ and helicity 2, hence they should be related to conformal primary gravitons of the same dimension. An expansion of the form \eqref{eq:stress-tensor-exp} has been applied to the news tensor in 4d asymptotically flat spacetimes (AFS) in \cite{Freidel:2022skz}, where it was shown that the coefficients with integer negative or positive $\mathfrak{sl}(2,\mathbb{C})$ dimension form a complete basis under certain conditions. 

Both eq. \eqref{eq:stress-tensor-exp} and our strategy to verify it are in principle very simple. We would like to start with the stress tensor contribution to the CFT$_3$ stress tensor OPE block 
\begin{equation}
\label{eq:stress-tensor}
    T^+(x_1) T^+(x_2) \supset c_{T^+T^+\widetilde{T}^-} \int d^3x_3 \langle   T^+(x_1) T^+(x_2) \widetilde{T}^-(x_3)\rangle T^+(x_3) + \cdots
\end{equation}
and show that\footnote{Upon accounting for an appropriate normalization involving powers of $c$, see Section \ref{sec:Sl2-from-so32}.}
\begin{equation}
\label{eq:3dblock-2dblock-intro}
   \oint du_1 u_1^{s_1} \oint du_2 u_2^{s_1}  \lim_{c \rightarrow 0} T^+(cu_1, \vec{z}_1) T^+(cu_2, \vec{z}_2) \questionequal \mathcal{G}^+_{2 - s_1}(\vec{z}_1)\mathcal{G}^+_{2 - s_2}(\vec{z}_2).
\end{equation}
For $s_1, s_2 \geq 0$, the RHS of eq. \eqref{eq:3dblock-2dblock-intro} should then agree with the $\mathfrak{sl}(2,\mathbb{C})$ block of conformally soft gravitons from which a $w_{1+\infty}$ algebra can be subsequently extracted using the same approach in \cite{Guevara:2021abz}. We now explain why establishing eq. \eqref{eq:3dblock-2dblock-intro} turned out to be harder than originally anticipated, nevertheless possible. 

Firstly, correlation functions of spinning particles in higher-dimensional conformal field theories are in general not simple. In the case of three-point functions, the $\mathfrak{so}(3,2)$ covariant three-point structures contributing to a spinning correlator can be enumerated \cite{Costa:2011mg}. For spin-2 operators, there are 10 allowed structures. Conservation of the CFT$_3$ stress tensor further restricts the linear combination of these structures to depend only on two free parameters. The final answer can be expressed in terms of weight-shifting operators acting on a scalar correlator (explicit formulas involving the stress tensor in CFT$_3$ can be found, for example, in Appendix A of \cite{Dymarsky:2017yzx}). We found that in the Carrollian limit, most of these structures are subleading. As a result, the leading contribution to the stress tensor 3-point function in \eqref{eq:stress-tensor} dramatically simplifies to a sum of only three terms, each of which can be expressed in terms of a simple weight-shifting operator acting on a scalar three-point function.  

The second problem is now clear: the Carrollian limits of scalar three-point functions have been computed in \cite{PipolodeGioia:2022exe, Bagchi:2023fbj, deGioia:2024yne, Alday:2024yyj}, producing distributional correlators belonging to the so-called electric Carroll subsector \cite{Bagchi:2016bcd, deBoer:2023fnj} of the CFT$_3$. Moments of these correlators with respect to $u$ computed via the integrals in \eqref{eq:3dblock-2dblock-intro} correspond to celestial amplitudes which are non-vanishing only when all three-points on the celestial sphere/plane are coincident \cite{Pasterski:2017ylz}. On the other hand, the algebra of conformally soft gravitons was computed using a standard CFT$_2$ OPE block. In order to obtain such correlators, we had to understand how to access so-called magnetic Carroll subsector \cite{Bagchi:2016bcd} of the CFT$_3$ in the limit. The difficulty now is that merely expanding the CFT$_3$ three-point functions in powers of $u_i$ cannot possibly yield 2d CFT correlators of negative dimension modes. This is because for fixed $\vec{z}_i$, the asymptotics of the CFT$_3$ correlators imply that the small-$u$ expansions in \eqref{eq:stress-tensor-exp} always start at $n = 0$ and hence only capture the $\mathfrak{sl}(2,\mathbb{C})$ modes of positive scaling dimension. 
We overcome this problem by using an integral representation of the CFT$_3$ correlator from which we can extract the whole tower of negative dimension modes. 

{\bf Summary of results.} In Section 4 we show that the $\mathfrak{so}(3,2)$ OPE block of scalars $\Ocft{\dcft}$ with dimensions $\dcft_1, \dcft_2$ can be decomposed into a collection of $\mathfrak{sl}(2,\mathbb{C})$ OPE blocks of operators with dimensions $\dccft_1, \dccft_2$ with $\dccft_i = \dcft_i - s_i - 1$ and $s_i \in \mathbb{Z}$. Discarding the distributional terms in $\vec{z}_{ij}$,  we find that, to leading order in the Carrollian limit, there is only one operator  of dimension $\dccft_1 + \dccft_2 - 2$ exchanged in each of these blocks. This block agrees with the celestial OPE block of scalars implied by collinear factorization of scattering amplitudes in 4d flat space \cite{Fan:2019emx, Pate:2019lpp, Himwich:2021dau}. 

In Section 5 we decompose the $\mathfrak{so}(3,2)$ OPE blocks of a conserved current ($\dcft_1 = 2, \ell_1 = 1$) and the stress tensor  ($\dcft_1 = 3, \ell_1 = 2$) respectively and a scalar $(\dcft_2 = \dcft, \ell_2 = 0)$ into a collection of  $\mathfrak{sl}(2,\mathbb{C})$ OPE blocks. We focus on the exchange of a scalar in the block involving transverse $(z)$ components of the spinning operators. There is a unique conformal structure that determines the associated three-point functions \cite{Costa:2011mg} which can be expressed in terms of a weight-shifting operator acting on the scalar three-point function. We use the results of Section 4 to show that, to leading order in the Carrollian limit, the $\mathfrak{sl}(2,\mathbb{C})$ blocks of the current and stress tensor modes with respective dimensions $1 - s$ and $2 - s$, $s \in \mathbb{N}$ reproduce the towers of tree-level soft gluon and graviton theorems in a conformal primary basis \cite{Hamada:2018vrw}.  

In Section 6 we analyze the $\mathfrak{so}(3,2)$ OPE blocks of conserved currents and stress tensors. We focus on exchanges of a conserved current and the stress tensor and analyze their transverse components. We express the associated three-point functions in terms of weight-shifting operators acting on scalar correlators. To leading order in the Carrollian limit, we obtain $\mathfrak{sl}(2,\mathbb{C})$ blocks of current modes of dimensions $\dccft_1 = 1 - s_1, \dccft_2 = 1 - s_2$ which precisely reproduce the $S$ algebra first found in celestial CFT for $s_1, s_2 \in \mathbb{N}$ \cite{Guevara:2021abz, Strominger:2021lvk}. Similarly, the $\mathfrak{sl}(2,\mathbb{C})$ blocks of stress tensor modes of dimensions $\dccft_1 = 2 - s_1, \dccft_2 = 2 - s_2$, $s_1, s_2 \in \mathbb{N}$ reproduce the $w_{1 + \infty}$ algebra. 
A $w_{1 + \infty}$ algebra in the Carrollian context was also discussed in \cite{Saha:2023hsl,Saha:2023abr}, however the analysis therein relied on the existence of a spin-2 operator in the Carrollian field theory. We emphasize that such an assumption is not needed here. We instead start with a CFT$_3$ (which contains a stress tensor by definition) and show that a spin-2 operator arises naturally as one of the modes in its $\mathfrak{sl}(2,\mathbb{C})$ mode decomposition, after taking a Carrollian limit. 

Our results clarify the sense in which the towers of 4d soft theorems leading to the S- and $w$-algebras are universal: they are a direct consequence of the decomposition of (a sector of) the unitary $\mathfrak{so}(3,2)$ representations $(\dcft,\ell) = (2,1),~(\dcft,\ell) = (3,2)$ into integer dimension $\mathfrak{sl}(2,\mathbb{C})$ representations.\footnote{General decompositions of $\mathfrak{so}(d,2)$ conformal blocks into $\mathfrak{so}(d,1)$ blocks were worked out in \cite{Hogervorst:2016hal}.} This result follows from the fact that the spin-$\ell$-scalar-scalar three-point function in CFT$_3$ is uniquely determined up to a coefficient. On the other hand, the towers of soft theorems are well-known to receive interesting and not yet fully understood corrections. We expect these corrections to be encoded in contributions from other operators exchanged in the CFT$_3$ current-scalar and stress tensor-scalar OPE blocks. These may be accessed in higher-point spinning correlators, which one can hope to determine in the future using bootstrap techniques for higher-dimensional CFTs.

Similarly, the $S$ and $w$ algebras are universal in the sense that both can be extracted from Carrollian expansions of components of the current and stress tensor three-point functions in CFT$_3$. For holographic CFTs, these are dual to gauge fields and gravitons in AdS$_4$. These algebras will receive corrections depending on the matter content of the bulk/boundary theories. Terms subleading in the Carrollian limit will also modify these algebras. Our work demonstrates that future progress in bootstrapping higher-dimensional CFTs, and in particular the CFT stress tensor 4-point function in \cite{Dymarsky:2017yzx,Chang:2024whx} will inevitably shed light on non-perturbative scattering amplitudes and, in particular, the scattering of 4 gravitons (or conformal primary gravitons) in 4d flat space \cite{Chowdhury:2019kaq}.

\section{Review: 2d conformal primaries from 3d CFT}
\label{sec:Sl2-from-so32}

Conformal primary scalar operators $\Ocft{\dcft}(x)$ in $d$-dimensional conformal field theories (CFT$_d$) transform under the $\mathfrak{so}(d,2)$ conformal algebra according to \cite{Osborn2019:lectures}
\begin{equation}
\label{eq:cft-gens}
\begin{split}
[T_{\mu}, \Ocft{\dcft}(x)] &= -i \p_{\mu} \Ocft{\dcft}(x), \\
[K_{\mu}, \Ocft{\dcft}(x)] &= - i \left( x^2 \p_{\mu} - 2 x_{\mu} x^{\nu}\p_{\nu} - 2 x_{\mu} \dcft \right) \Ocft{\dcft}(x),\\
[D,\Ocft{\dcft}(x)] &= -i\left(x^{\mu}\p_{\mu} + \dcft \right) \Ocft{\dcft}(x),\\
[J_{\mu\nu}, \Ocft{\dcft}(x)] &= -i \left( x_{\mu} \p_{\nu} - x_{\nu} \p_{\mu} \right)\Ocft{\dcft}(x).
\end{split}
\end{equation}
Here, $T_{\mu}, K_{\mu}, D$ and $J_{\mu\nu}$ are generators of $\mathfrak{so}(d,2)$. They correspond respectively to translations, special conformal transformations, dilations and boosts in the CFT$_d$. From now on, we will set $d = 3$. 
The $\mathfrak{so}(3,2)$ algebra is related to the 4d Poincar\'e algebra by a well-known In\"on\"u-Wigner (IW) contraction \cite{Inonu}. This can be seen by introducing coordinates $(t, z, \bar{z})$ on the plane, rescaling 
\begin{equation}
\label{eq:Carroll-contraction}
    t =  c u,\quad 
    \mathcal{T}_0 + \mathcal{T}_3 = c T_t, \quad \mathcal{T}_0 - \mathcal{T}_3 = c K_t, \quad \mathcal{T}_1 = c\left(J_{tz} + J_{t\bz} \right), \quad \mathcal{T}_2 = -ic(J_{tz} - J_{t\bz}),
\end{equation}
and taking $c \rightarrow 0$ for fixed $u$ and $\mathcal{T}_i$. The  3d formulas for the generators in eq. \eqref{eq:cft-gens} are given in Appendix \ref{app:3d-algebra}, from which one can easily verify that the IW contraction yields asymptotic 4d Poincar\'e generators. These kinematic considerations alone suggest that flat space observables such as scattering amplitudes can be obtained from a limit of conformal correlators in CFT$_3$. Indeed, the flat space limit has led to many recent insights into perturbative and non-perturbative aspects of scattering amplitudes \cite{Gary:2009mi, Penedones:2011writing,Paulos:2016fap,Lipstein:2019mpu,Jain:2022ujj,Marotta:2024sce}.

In fact more is true. The allowed space of solutions to the conformal Killing equations in 3d in the Carrollian limit 
\begin{equation}
\label{eq:Carr-lim}
    t = cu, \quad {\rm with} ~~c \rightarrow 0, \quad  u ~~ {\rm fixed}
\end{equation}
is infinitely dimensional. The contraction \eqref{eq:Carroll-contraction} applied to this enhanced algebra then generates the extended BMS algebra in 4d \cite{Duval:2014uva}. Furthermore, the leading and subleading conformally soft graviton Ward identities arise directly from the Ward identities of the stress tensor in CFT$_3$ \cite{de2023celestial}. To see this, one first computes BMS$_4$ primaries from CFT$_3$ by taking the Carrollian limit \cite{PipolodeGioia:2022exe,Bagchi:2023fbj}
\begin{equation}
\label{eq:carroll}
    \Ocarr{\dcft}(u,z,\bz) = \lim_{c \rightarrow 0} i^{\Delta} c^{\Delta - 1} \Gamma\left(\Delta - \frac{1}{2}\right) \Ocft{\dcft}(cu, z, \bz),
\end{equation}
where the $c\to 0$ limit should be understood in the weak sense, i.e., after forming a correlator out of a set of operators. We choose the normalization of $\Ocft{\dcft}$ to be inherited from the AdS$_4$ bulk-to-boundary propagator, which in terms of embedding space variables takes the form \cite{Costa:2014spinning}
\begin{equation}
    K_{\Delta}(P; X) =  \frac{2^{\dcft}\Gamma(\dcft)}{2\pi^{3/2}\Gamma(\dcft - \frac{1}{2})} \frac{1}{(-2P\cdot X)^{\dcft}}.
\end{equation}
In the Carrollian limit, the associated two-point function will diverge as $c^{2(1 - \dcft)}$. The power of $c$ in \eqref{eq:carroll} is chosen to cancel this divergence. We will also need the generalization of \eqref{eq:carroll} for spinning operators
\begin{equation}
    \label{eq:carroll-spin}
    \Ocarr{\dcft}^{\ell,\mu_1 \cdots \mu_{\ell}}(u,z,\bz) = \lim_{c \rightarrow 0} i^{\Delta} c^{\Delta + \ell - 1}\frac{\dcft - 1}{\ell + \dcft - 1}\Gamma\left(\Delta - \frac{1}{2}\right) \Ocft{\dcft}^{\ell,\mu_1\cdots \mu_{\ell}}(cu, z, \bz).
\end{equation}
The normalization can again be inferred from the flat space limit of the associated AdS bulk-to-boundary propagator\footnote{Our conventions are related to those in \cite{Costa:2014spinning} by $P_{\rm here} = P_{\rm there}/2$} \cite{Costa:2014spinning}
\begin{equation}
    K^{\ell}_{\dcft}(P, Z;X, W) = \frac{2^{\dcft + \ell}(\ell + \dcft -1)\Gamma(\dcft)}{2\pi^{3/2}(\dcft - 1)\Gamma(\dcft -\frac{1}{2})} \frac{(-(P\cdot X) (W\cdot Z) + ( W\cdot P) (Z\cdot X))^{\ell}}{(-2P\cdot X)^{\dcft + \ell}}.
\end{equation}

$\Ocarr{\dcft}(u,z,\bz)$ can be transformed to a 2d conformal primary basis by evaluating the integral \cite{Donnay:2022aba} 
\begin{equation}
    \label{eq:int-tr}
\Occft{\delta}(z, \bar{z}) \equiv  N(\Delta, \delta)   \int_{-\infty}^{\infty} du (u + i\eta\epsilon)^{\Delta - \delta - 1}  \Ocarr{\Delta}( u, z, \bar{z}).
\end{equation}
The normalization $N(\Delta, \delta)$ given by \cite{de2023celestial} 
\be 
\label{eq:norm}
N(\Delta, \delta) =  \pi^{\frac{1}{2}}(\eta i)^{\delta + 1 - \Delta} \Gamma(\delta + 1 - \Delta)
\ee
and is chosen such that the operators on the LHS of eq. \eqref{eq:int-tr} agree with the celestial operators obtained through a Mellin transform of 4d plane waves.\footnote{We take the scalar conformal primary wavefunctions to be normalized as $\varphi_{\dccft}(q;x) = \frac{i^{\dccft}\Gamma(\dccft)}{(-q\cdot x)^{\dccft}}$, while the spinning wavefunctions are obtained by dressing these with the corresponding polarization tensors \cite{Pasterski:2020pdk}.} Here $\eta = \pm 1$ depending on whether the mode is outgoing or incoming. We would like to emphasize that the integral transform \eqref{eq:int-tr} does not commute with the Carrollian limit (this will be elaborated on in \cite{Navarro:2025}). It is therefore important to take the Carrollian limit of CFT correlators prior to computing the integral transforms \eqref{eq:int-tr}. When applied to CFT$_3$ 3- and 4-point functions, this procedure was shown to yield correctly normalized celestial amplitudes in either Euclidean or Lorentzian signature \cite{deGioia:2024yne,Alday:2024yyj,Lipstein:2025jfj,Kulkarni:2025qcx}.

 It is straightforward to check that the operators defined in eq. \eqref{eq:carroll} obey the conformal Carroll Ward identities (see \cite{Bagchi:2025vri} for a review) with boost dimension $\Delta$.\footnote{To see this, note that the $\mathfrak{so}(3,2)$ generators in \eqref{eq:3d-so32} reduce to the global $\mathfrak{ccarr}(3)$ generators in eqns. (5.1a) and (5.1b) of \cite{Bagchi:2025vri} in the Carrollian limit. Thereby, the 3d conformal Ward identities will become Carrollian Ward identities upon IW contraction.}
Equivalently, the operators defined in eq. \eqref{eq:int-tr} are 2d conformal primaries at $(z,\bar{z}) = (0,0)$. This can be seen from eq. \eqref{eq:3d-so32} in Appendix \ref{app:3d-algebra} which implies
\begin{equation}
   -i \lim_{c \rightarrow 0} K_{\bz}(cu, z, \bz) = \mathcal{L}_1, \quad -\lim_{c \rightarrow 0}i K_{z}(cu, z, \bz) =\bar{\mathcal{L}}_1.
\end{equation}
Together with $iT_{z}, iT_{\bz} ~(\mathcal{L}_{-1}, \bar{\mathcal{L}}_{-1}), iD \equiv \mathcal{L}_0 + \bar{\mathcal{L}}_0$ and $2iJ_{z\bz} \equiv -\mathcal{L}_0 + \bar{\mathcal{L}}_0$, these generate an $\mathfrak{sl}(2,\mathbb{C})$ algebra
\be 
\begin{split}
[\mathcal{L}_m ,\mathcal{L}_n] &= (n - m)\mathcal{L}_{m+n},  \\
[\bar{\mathcal{L}}_m ,\bar{\mathcal{L}}_n] &= (n - m)\bar{\mathcal{L}}_{m+n}, \quad m, n = -1, 0, 1.
\end{split}
\ee

\section{Integer sl$(2,\mathbb{C})$ primaries from so$(3,2)$ primaries}
\label{sec:hierarchy}

It was shown in \cite{Freidel:2022skz} that 2d conformal primaries of respectively positive and negative integer scaling dimensions form a \textit{complete} basis for functions on $\mathbb{R}\times S^2$ belonging to the Schwartz space. Related ideas were discussed in \cite{Cotler:2023qwh}. 
In this section, we apply this decomposition to $\mathfrak{so}(3,2)$ conformal primaries.\footnote{In general, primary operators in CFT$_3$ do not belong to the Schwartz space. Their large-distance asymptotics can be read from their insertions into correlation functions
\begin{equation}
\label{eq:primary-asy}
  \langle \Ocft{\dcft}(x) \cdots \rangle  \sim  \frac{1}{x^{2\Delta}}, \quad  x^2 \rightarrow \infty.
\end{equation}
At large $t$ and fixed $\vec{z}$, $x^2 \rightarrow -t^2$ and so \eqref{eq:primary-asy} clearly violates the Schwartz condition. 
We can remedy \eqref{eq:primary-asy} by introducing a regulator $\langle \Ocft{\dcft}(x) \cdots \rangle \rightarrow \langle \Ocft{\dcft}(x) \cdots \rangle e^{\epsilon x^2}$. Whether the integer modes form a complete basis or not will not be relevant here, although it will be important in establishing a 1-1 correspondence between CFT$_3$ and families of celestial correlators.}

We consider the Laurent expansion in $t$ of an $\mathfrak{so}(3,2)$ primary
\begin{equation}
\label{eq:Carroll-exp1}
  \Ocft{\dcft}(t, \vec{z}) = \sum_{n = 0}^{\infty} t^n \cftm{\dcft}{n}(\vec{z}).
\end{equation}
That the leading term in this expansion is $\mathcal{O}(t^0)$ follows from analyzing the small $t$, fixed $\vec{z}$ limit of CFT$_3$ two- and three-point functions. Subtleties may arise when $|\vec{z}| \sim t$. In the following, we will always take $t \rightarrow 0$ first.  
It will also be useful to introduce the Fourier transform of $\Ocft{\dcft}$ with respect to $t$
\begin{equation}
\label{eq:FT-3d}
   \fOcft{\dcft}(\omega, \vec{z}) \equiv \int_{-\infty}^{\infty} dt e^{-i\omega t }  \Ocft{\dcft}(t, \vec{z}) = \sum_{n} \omega^n \fcftm{\dcft}{n}(\vec{z}).
\end{equation}
 We see that the modes of the CFT$_3$ primaries are related to modes of the Carrollian operators defined in eq. \eqref{eq:carroll}
\begin{equation}
\label{eq:Carr-mode}
    \Ocarr{\dcft}(u,z,\bz) = \sum_{n = 0}^{\infty} u^n \carrm{\dcft}{n}(z,\bz)
\end{equation}
via
\begin{equation}
\label{eq:carr-cft-p}
    \carrm{\dcft}{n}(z,\bz) = \lim_{c \rightarrow 0} i^{\Delta} c^{\Delta + n - 1} \Gamma\left(\Delta - \frac{1}{2}\right) \cftm{\dcft}{n}(\vec{z}).
\end{equation}
These Carrollian modes transform like $\mathfrak{sl}(2,\mathbb{C})$ primaries of dimensions $\Delta + n$. To see this, we invert \eqref{eq:Carr-mode}
\begin{equation}
\label{eq:Carrp}
    \carrm{\dcft}{n}(\vec{z}) =  \oint \frac{du}{2\pi i} u^{-n-1} \Ocarr{\dcft}(u, \vec{z}) = \frac{1}{n!} \lim_{u \rightarrow 0}\p_u^n \Ocarr{\dcft}(u, \vec{z})
\end{equation}
which obey
\begin{equation}
\label{eq:action}
\begin{split}
    [\mathcal{L}_1, \carrm{\dcft}{n}(\vec{z})] &\equiv -[\lim_{c \rightarrow 0} iK_{\bz}(cu, \vec{z}), \carrm{\dcft}{n}(\vec{z})] =  \left(z^2 \p_z + z(\Delta + n) \right)\carrm{\dcft}{n}(\vec{z}), \\
    [\mathcal{L}_0 + \bar{\mathcal{L}}_0,\carrm{\dcft}{n}(\vec{z})] &\equiv [iD,  \carrm{\dcft}{n}(\vec{z})] = (\Delta + n +z\p_z +\bz \p_{\bz})\carrm{\dcft}{n}(\vec{z}),\\
     [\mathcal{L}_0 - \bar{\mathcal{L}}_0, \carrm{\dcft}{n}(\vec{z})] &\equiv -2 [iJ_{z\bz}, \carrm{\dcft}{n}(\vec{z})] = (z\p_z - \bz \p_{\bz})\carrm{\dcft}{n}(\vec{z}), \\
     [\mathcal{L}_{-1}, \carrm{\dcft}{n}(\vec{z})] & \equiv [iT_z,\carrm{\dcft}{n}(\vec{z})] =   \p_z\carrm{\dcft}{n}(\vec{z}).
    \end{split}
\end{equation}

Similarly, the modes of the Fourier transform \eqref{eq:FT-3d} are related to Mellin modes of the Fourier transform  of \eqref{eq:Carr-mode}
\begin{equation}
\label{eq:carrft}
   \Ocarrft{\dcft}(\omega,z,\bz) \equiv \int_{-\infty}^{\infty} du e^{-i\omega u } \Ocarr{\dcft}(u,z,\bz) =  \sum_{n = 0}^{\infty} \omega^{n} \fcarrm{\dcft}{n}(\vec{z})
\end{equation}
via
\begin{equation}
\label{eq:carr-cft-m}
     \fcarrm{\dcft}{n}(\vec{z}) = \lim_{c \rightarrow 0} i^{\Delta} c^{\Delta -n-2} \Gamma\left(\Delta - \frac{1}{2} \right) \fcftm{\dcft}{n}(\vec{z}).
\end{equation} 
Eq. \eqref{eq:carrft} can be inverted to express $\fcarrm{\dcft}{n}$ in terms of $\Ocarrft{\dcft}$ namely,
\begin{equation}
\label{eq:FM}
\begin{split}
    \fcarrm{\dcft}{n}(\vec{z}) &= \oint \frac{d\omega}{2\pi i} \omega^{-n-1} \Ocarrft{\dcft}(\omega, \vec{z}) = \frac{1}{n!}\lim_{\omega \rightarrow 0}\p_{\omega}^{n} \Ocarrft{\dcft}(\omega, \vec{z}) \\
     &=  \frac{(-i)^n}{n!} \int_{-\infty}^{\infty} du (u + i\epsilon)^{n}  \Ocarr{\dcft}(u, \vec{z}).
     \end{split}
\end{equation}
As a result, the operators in \eqref{eq:FM} transform as $\mathfrak{sl}(2,\mathbb{C})$ primaries of dimension $\Delta - n - 1$, 
\begin{equation}
\label{eq:fourier}
    \begin{split}
    [\mathcal{L}_1,  \fcarrm{\dcft}{n}(\vec{z})] & =  \left(z^2 \p_z + z(\Delta - n -1) \right)  \fcarrm{\dcft}{n}(\vec{z}), \\
    [\mathcal{L}_0 + \bar{\mathcal{L}}_0,  \fcarrm{\dcft}{n}(\vec{z})] &= (\Delta - n - 1 +z\p_z +\bz \p_{\bz}) \fcarrm{\dcft}{n}(\vec{z}),\\
     [\mathcal{L}_0 - \bar{\mathcal{L}}_0,  \fcarrm{\dcft}{n}(\vec{z})] &= (z\p_z - \bz \p_{\bz}) \fcarrm{\dcft}{n}(\vec{z}), \\
     [\mathcal{L}_{-1},  \fcarrm{\dcft}{n}(\vec{z})] & =   \p_z  \fcarrm{\dcft}{n}(\vec{z}).
    \end{split}
\end{equation}
The action of the translation generators obtained by IW contraction of the remaining $\mathfrak{so}(3,2)$ generators can be obtained similarly. 

Combining \eqref{eq:Carrp} and \eqref{eq:FM} with \eqref{eq:int-tr}, \eqref{eq:carr-cft-p} and \eqref{eq:carr-cft-m}, we conclude that an $\mathfrak{so}(3,2)$ primary of dimension $\dcft$ decomposes in the Carrollian limit into a tower of $\mathfrak{sl}(2,\mathbb{C})$ modes 
\begin{equation}
\label{eq:ccft-carr}
\begin{split}
    \Occft{\dcft+n}(\vec{z}) &= 2\pi i N(\dcft, \dcft + n) \carrm{\dcft}{n}(\vec{z}), \\
     \Occft{\dcft - n - 1}(\vec{z}) &= i^n  n! N(\dcft, \dcft - n - 1) \fcarrm{\dcft}{n}(\vec{z}). \\
    \end{split}
\end{equation}
The factor $N$ given in \eqref{eq:norm} identifies these modes with properly normalized celestial CFT operators. In summary, both sets of $\mathfrak{sl}(2,\mathbb{C})$ modes can be obtained from an $\mathfrak{so}(3,2)$ primary by first applying \eqref{eq:carroll} then performing the integral transform \eqref{eq:int-tr}.

We emphasize that in extracting flat space observables from CFT$_3$ ones, it is crucial to first take the $c \rightarrow 0$ limit and only then select particular $2d$ modes by applying integral transforms such as \eqref{eq:Carrp} and \eqref{eq:FM}. In the absence of the limit, the different $\mathfrak{sl}(2,\mathbb{C})$ subsectors in the expansion \eqref{eq:Carroll-exp1} mix under the $\mathfrak{so}(3,2)$ action. This can be seen by considering the action of the special conformal 
generator on a mode in the expansion \eqref{eq:Carroll-exp1}
\begin{equation}
\label{eq:cft3}
[ K_{\bz}(t, \vec{z}),  \cftm{\dcft}{n}(\vec{z})]= \oint \frac{dt}{2\pi i} t^{-n - 1} [ K_{\bz}(t, \vec{z}),  \Ocft{\dcft}(t, \vec{z})] \propto [\mathcal{L}_1, \cftm{\dcft}{n}(\vec{z})] + \p_{\bz}  \cftm{\dcft}{n-2}(\vec{z}).
\end{equation}
On the other hand, taking the Carrollian limit while accounting for the normalization \eqref{eq:carr-cft-p} allows us to project the $\mathfrak{so}(3,2)$ primary in \eqref{eq:Carroll-exp1} onto a particular $\mathfrak{sl}(2,\mathbb{C})$ subsector. In this case, the second term in \eqref{eq:cft3} becomes suppressed by a factor of $c^2$ with respect to the first and drops out in the $c \rightarrow 0$ limit, namely  
\begin{equation}
  [K_{\bz}(t, \vec{z}), \carrm{\dcft}{n}(\vec{z})] \propto  \lim_{c \rightarrow 0} c^{\Delta + n - 1} \oint \frac{dt}{2\pi i} t^{-n - 1} [ K_{\bz}(t, \vec{z}),  \Ocft{\dcft}(t, \vec{z})] \propto [\mathcal{L}_1, \carrm{\dcft}{n}(\vec{z})].
\end{equation}

In the next section we will apply these formulas to decompose the correlation function of scalar $\mathfrak{so}(3,2)$ primaries into correlators of $\mathfrak{sl}(2,\mathbb{C})$ modes to leading order in the Carrollian limit. This will allow us to extract an integer hierarchy of $\mathfrak{sl}(2,\mathbb{C})$ blocks from the 3d OPE block \cite{Ferrara,Ferrara:1973eg} of scalar primaries. We will then express the spinning OPE blocks in terms of weight-shifting operators acting on the scalar blocks. This will allows us to show in Sections \ref{sec:soft-tower} and \ref{eq:higher-spin} that the towers of soft gluon and graviton theorems in 4d flat space \cite{Hamada:2018vrw}, as well as the associated infinite towers of symmetry algebras \cite{Guevara:2021abz, Strominger:2021lvk} appear, to leading order in the Carrollian limit, as a result of the decomposition of the $\mathfrak{so}(3,2)$ OPE blocks of currents and the stress tensor into $\mathfrak{sl}(2,\mathbb{C})$ subsectors.

\section{Dimensional reduction of CFT$_3$ scalar OPE block}
\label{sec:dim-red-scalar}

Consider the OPE block of scalar CFT$_3$ primaries \cite{Ferrara,Ferrara:1973eg}
\begin{equation}
\label{eq:scalar-block}
    \Ocft{\dcft_1}(t_1, \vec{z}_1)   \Ocft{\dcft_2}(t_2, \vec{z}_2) \supset \mathcal{N}_S \int dt_3 \int d^2 \vec{z}_3 \langle   \Ocft{\dcft_1}(t_1, \vec{z}_1)   \Ocft{\dcft_2}(t_2, \vec{z}_2)   \Ocft{3 - \dcft_3}^S(t_3, \vec{z}_3) \rangle \Ocft{\dcft_3}(t_3, \vec{z}_3).
\end{equation}
Here $\Ocft{3 - \dcft_3}^S$ is the 3d shadow transform of $\Ocft{\dcft_3}$ defined by
\begin{equation}
\label{eq:shadow}
  \Ocft{3 - \dcft_3}^S(P_3) \equiv \int d^3P \frac{1}{(-2P_3 \cdot P)^{3 - \dcft_3}}  \Ocft{\dcft_3}(P)
\end{equation}
and
\begin{equation}
\label{eq:shadow-norm}
\mathcal{N}_S = \frac{\Gamma(\dcft_3) \Gamma(3 - \dcft_3)}{\pi^3 \Gamma(\dcft_3 - \frac{3}{2})\Gamma(\frac{3}{2} - \dcft_3)}
\end{equation}
is a normalization that schematically ensures that $\int \Ocft{3 - \Delta_3}^S \Ocft{\Delta_3}$ acts as the identity operator \cite{SimmonsDuffin:2012uy}. 

The goal of this section is to extract correlation functions of $\mathfrak{sl}(2,\mathbb{C})$ primaries of arbitrary dimension from the three-point functions of scalar $\mathfrak{so}(3,2)$ primaries and consequently  dimensionally reduce the $\mathfrak{so}(3,2)$ OPE block \eqref{eq:scalar-block} into a collection of $\mathfrak{sl}(2,\mathbb{C})$ blocks. Our strategy was already outlined in Section \ref{sec:hierarchy}: we first need to compute the Carrollian limit accounting for the normalization in eq. \eqref{eq:carroll} and then evaluate the integral transforms in eq. \eqref{eq:int-tr}. We will be  interested in the $\mathfrak{sl}(2,\mathbb{C})$ blocks for the negative integer modes \eqref{eq:carr-cft-m}, but the same procedure can be used to obtain $\mathfrak{sl}(2,\mathbb{C})$ blocks of \textit{arbitrary}, and in particular, positive integer dimension modes.

In computing the Carrollian limit of the scalar CFT$_3$ three-point function, we immediately run into a naive difficulty: the $c \rightarrow 0$ limit results in correlators involving delta function contact terms in $\vec{z}_{ij}$. These are known to characterize the electric Carroll sector \cite{Bagchi:2016bcd}. They are also known to reproduce the expected celestial amplitudes dual to scattering amplitudes in 4d flat space which are always singular due to bulk momentum conservation. This limit gives rise to a degenerate 2d OPE block, unlike the OPE blocks governing the universal dynamics of conformal primaries in celestial CFT, or equivalently the collinear limits of massless particles in 4d flat space \cite{Fan:2019emx, Pate:2019lpp}. 

As shown in \cite{deGioia:2024yne}, the expansion of the CFT$_3$ three point function in the $c \rightarrow 0$ limit also involves a series of terms with power law dependence in $|\vec{z}_{ij}|$ as characteristic for standard 2d CFT (these correspond to observables in the so-called magnetic Carroll subsector of the CFT$_3$ \cite{Bagchi:2025vri}). We are interested in the projection of the 3d OPE block to the magnetic Carroll subsector because we would like to recover the standard 2d CFT OPE block of primary operators used in \cite{Guevara:2021abz} to compute the algebra of conformally soft gluons and gravitons. It is not immediately clear how to obtain a correlator of magnetic branch modes from an expansion of the CFT$_3$ 3-point function at small $t_i$ since this expansion will naively have no poles in $t_i$ -- see eq. \eqref{eq:Carroll-exp1}. 

In this section, we show that the correlators of modes of \textit{any} dimension in the magnetic Carroll subsector of a CFT$_3$ can, in fact, be computed from limits of CFT$_3$ 3-point correlators by using an integral representation thereof. Our analysis will confirm the suggestion of \cite{Bagchi:2022emh,Banerjee:2024hvb} that this sector governs the dynamics of soft particles in 4d flat space, provided that one restricts to $\mathfrak{sl}(2,\mathbb{C})$ primaries of negative integer dimensions.\footnote{Unlike in \cite{Banerjee:2024hvb}, we do not need to consider further integral transforms, such as the light transform in the 2d CFT.} This restriction is however not necessary: our analysis will allow us to project a CFT$_3$ three-point function to the magnetic Carroll branch from which one can obtain standard CFT$_2$ correlators of arbitrary dimension.

We consider the time-ordered scalar CFT$_3$ 3-point function on the Lorentzian plane 
\begin{equation}
\label{eq:plane-3-pt}
    \langle \Ocft{\dcft_1}(x_1)\Ocft{\dcft_2}(x_2)\Ocft{\dcft_3}(x_3)  \rangle = \frac{c_{123}}{\left(-t_{12}^2 + |z_{12}|^2 + i\epsilon\right)^{\beta_{12}}\left(-t_{13}^2 + |z_{13}|^2 + i\epsilon \right)^{\beta_{13}}\left(-t_{23}^2 + |z_{23}|^2 + i\epsilon \right)^{\beta_{23}}} .
\end{equation}
This is related to the Carrollian limit of 3-point functions on the Lorentzian cylinder with operator insertions in the vicinity of the same global time slice, corresponding to an all-outgoing particle configuration from the bulk point of view.
All other in/out configurations can be obtained multiplying eq. \ref{eq:plane-3-pt} by an appropriate phase \cite{deGioia:2024yne}. 

We first express \eqref{eq:plane-3-pt} as
\begin{equation}
\label{eq:master-f}
\begin{split}
 \langle \Ocft{\dcft_1}(x_1)\Ocft{\dcft_2}(x_2)\Ocft{\dcft_3}(x_3)  \rangle &= c_{123} \frac{i^{-\frac{\Delta_1 + \Delta_2 + \Delta_3 + 1}{2}}}{\Gamma(\beta_{12})\Gamma(\beta_{13})\Gamma(\beta_{23})}  \prod_{i = 1}^3 \left( \int_0^{\infty} d\alpha_i  \alpha_i^{\Delta_i - 1} \right) \alpha^{-\frac{\Delta_1 + \Delta_2 + \Delta_3}{2}} \left(\frac{\alpha}{\pi} \right)^{3/2}  \\
&\times \int dt \int d^2z  e^{i\sum_{i = 1}^3 \alpha_i (-(t - t_i)^2 + |z - z_i|^2 + i\epsilon)},
\end{split}
\end{equation}
where 
\begin{equation}
\label{eq:alpha}
    \beta_{12} = \frac{\Delta_1 + \Delta_2 - \Delta_3}{2}, \quad  \beta_{13} = \frac{\Delta_1 + \Delta_2 - \Delta_3}{2}, \quad \beta_{23} = \frac{\Delta_2 + \Delta_3 - \Delta_1}{2}
\end{equation}
and
\begin{equation}
    \alpha \equiv \sum_{i=1}^3 \alpha_i.
\end{equation}
This representation is familiar from the computation of 3-point AdS$_{d+1}$ Witten diagrams \cite{witten1998anti}, but it holds irrespective of whether the CFT$_3$ admits a holographic dual or not. Eq. \eqref{eq:master-f} is obtained from eq. \eqref{eq:id0} derived in Appendix \ref{app:derivation} by setting $\alpha_i \rightarrow -i \alpha_i$.\footnote{Up to an overall factor of $i^2$ from the evaluation of Gaussian integrals over $t$ and $z, \bz$ respectively.} 

The simple, yet key, observation is that we can perform a Carrollian expansion of \eqref{eq:plane-3-pt} by setting $t_i = c u_i$ in \eqref{eq:master-f} and noticing that to leading order in the small $c$ expansion, we can approximate
\begin{equation}
-(t - t_i)^2 = -t (t - 2c u_i) + \mathcal{O}(c^2u_i^2).
\end{equation}
This allows us to write
\begin{equation}
\label{eq:3d-leadingc}
\begin{split}
\langle \Ocft{\dcft_1}(x_1)\Ocft{\dcft_2}(x_2)\Ocft{\dcft_3}(x_3)  \rangle &=  c_{123}\frac{i^{-\frac{\Delta_1 + \Delta_2 + \Delta_3 +1}{2}}}{\Gamma(\beta_{12})\Gamma(\beta_{13})\Gamma(\beta_{23})} \pi^{-3/2} \prod_{i = 1}^3 \left( \int_0^{\infty} d\alpha_i  \alpha_i^{\Delta_i - 1} \right) \alpha^{-\frac{\Delta_1 + \Delta_2 + \Delta_3-3}{2}}   \\
&\times \int dt \int d^2z  e^{i\sum_{i = 1}^3 \alpha_i (-t^2 + 2cu_i t + |z - z_i|^2 + i\epsilon)}\left(1 + \mathcal{O}(c^2 u_i^2) \right).
\end{split}
\end{equation}

We can now obtain from \eqref{eq:3d-leadingc} a correlator of Carrollian operators by applying \eqref{eq:carroll}. From these one computes correlators of $\mathfrak{sl}(2,\mathbb{C})$ (or celestial) modes by evaluating the integral transforms \eqref{eq:int-tr} with $\eta_i = 1$. For the negative dimension modes in eq. \eqref{eq:ccft-carr} (i.e. those obtained from \eqref{eq:int-tr} with $\dcft - \dccft - 1 = s \geq 0$), we find 
\begin{equation}
\label{eq:2d-corr}
\begin{split}
\langle \Occft{\Delta_1 - s_1-1}(\vec{z}_1) \Occft{\Delta_2 - s_2-1}(\vec{z}_2) \Occft{\Delta_3 - s_3 -1}(\vec{z}_3) \rangle 
&= \lim_{c\rightarrow 0} c^{\sum_i (\Delta_i - s_i - 1) -3}  \mathcal{N}_{\dcft_1\dcft_2\dcft_3} \prod_{i = 1}^3 \left( \int_0^{\infty} d\alpha_i  \alpha_i^{\Delta_i - s_i - 2} \right) \\
&\times \alpha^{ \frac{s_1 + s_2 + s_3 + 2}{2}}   \alpha^{-\frac{\Delta_1 + \Delta_2 + \Delta_3 -3}{2}} \int d^2z   e^{i\sum_{i = 1}^3 \alpha_i (|z - z_i|^2 - i\epsilon)},
\end{split}
\end{equation}
where we defined
\begin{equation}
\mathcal{N}_{\dcft_1\dcft_2\dcft_3} = \frac{i^{\frac{\sum_i \Delta_i- 1}{2}+3\frac{\sum_i s_i}{2} }}{\Gamma\left(\frac{s_1 + s_2 + s_3}{2} + 2\right)} \frac{\pi^4 c_{123} }{\Gamma(\beta_{12})\Gamma(\beta_{13})\Gamma(\beta_{23})}  2^{-\sum_i s_i} \prod_{i = 1}^3\Gamma\left(\Delta_i - \frac{1}{2}\right) .
\end{equation}
Here, $s_i$ should not be confused with helicities. They are integers that distinguish scalar $\mathfrak{sl}(2,\mathbb{C})$ primaries of different scaling dimensions. 

In obtaining this result, it was crucial to take the $c \rightarrow 0$ limit in \eqref{eq:3d-leadingc} prior to evaluating the $u_i$ integrals. It is not hard to see that otherwise,  the integrand will also contain a series of terms proportional to $c^{2n} u_i^{2n}$, all of which will contribute at the same order in $c$ after evaluating the $u_i$ integrals. Summing all these terms (or equivalently, evaluating the $t_i$ moments of the CFT$_3$ correlator without taking a Carrollian limit) should result in a 2d correlator that is fully $\mathfrak{so}(3,2)$ covariant.\footnote{We thank Andy Strominger for a discussion on this point.} In the case of the CFT$_3$ 2-point function, one can explicitly check that the Carrollian limit does not commute with integral transforms of the form \eqref{eq:int-tr}.

 In the case where the scalar CFT$_3$ correlator is dual to a contact diagram in AdS$_4$, the OPE coefficient $c_{123}$ is given by
\begin{equation}
\label{eq:scalar-AdS4}
c_{123} =  2^{\frac{\Delta_1 + \Delta_2 + \Delta_3}{2} - 3} g c^{-1} \frac{1}{2\pi^{3}}\frac{\Gamma(\beta_{12})\Gamma(\beta_{13})\Gamma(\beta_{23})\Gamma(\frac{\sum_i \Delta_i - 3}{2})}{\Gamma(\Delta_1 - \frac{1}{2})\Gamma(\Delta_2 - \frac{1}{2})\Gamma(\Delta_3 - \frac{1}{2})} .
\end{equation}
We finally apply \eqref{eq:id0} again, this time with $n = 2$ and
\begin{equation}
\Delta_i \rightarrow \delta_i \equiv \Delta_i - s_i - 1
\end{equation}
to recast \eqref{eq:2d-corr} into
\begin{equation}
\label{eq:2dccft}
\begin{split}
\langle \Occft{\dccft_1}(\vec{z}_1) &\Occft{\dccft_2 }(\vec{z}_2) \Occft{\dccft_3}(\vec{z}_3) \rangle =\frac{\mathcal{C}_{\dcft_1,\dcft_2,\dcft_3}(\delta_1, \delta_2, \delta_3) \lim_{c\rightarrow 0} c^{\sum_i \dccft_i -4}}{(|z_{12}|^2-i\epsilon)^{\frac{\dccft_1 + \dccft_2 - \dccft_3}{2}} (|z_{13}|^2-i\epsilon)^{\frac{\dccft_1 + \dccft_3 - \dccft_2}{2}} (|z_{23}|^2 -i\epsilon)^{\frac{\dccft_3 + \dccft_2 - \dccft_1}{2}} } ,
\end{split}
\end{equation}
where $\mathcal{C}_{\dcft_1\dcft_2\dcft_3}$ is given by
\begin{equation}
\label{eq:c123n}
\begin{split}
\mathcal{C}_{\dcft_1,\dcft_2,\dcft_3}(\delta_1, \delta_2, \delta_3) &=  g \pi^{2} 2^{\sum_i (\dccft_i -  \frac{\dcft_i}{2}) -1}  i^{\sum_i (\dcft_i+s_i) - 1} \frac{\Gamma(\frac{\delta_1 + \delta_2 - \delta_3}{2})\Gamma(\frac{\delta_1 + \delta_3 - \delta_2}{2})\Gamma(\frac{\delta_2 + \delta_3 - \delta_1}{2}) \Gamma\left(\frac{\sum_i \Delta_i - 3}{2}\right)}{\Gamma\left(\frac{s_1 + s_2 + s_3}{2} + 2\right)}.
\end{split}
\end{equation}

 Eq. \eqref{eq:2dccft} is the expected OPE block of 2d global conformal primaries of dimensions $\dccft_i$.  
 Using the identity\footnote{This identity technically only holds for $\lambda \in i \mathbb{R}$. Here we apply it in a generalized sense, see \cite{Donnay:2020guq} for an attempt to make this precise.}
 \begin{equation}
     \lim_{c \rightarrow 0}c^{-\nu} = 2\pi \nu \delta(\nu)
 \end{equation}
 we see that \eqref{eq:c123n} becomes proportional to a delta function in dimensions 
 \begin{equation}
 \label{eq:c-distr}
     \lim_{c \rightarrow 0} c^{\sum_i\delta_i - 4} = -2\pi \left(\sum_{i}\delta_i - 4\right)\delta\left(\sum_{i}\delta_i - 4 \right).
 \end{equation}
 This delta function is a reflection of 4d scale invariance. It will be important later on that, up to functions of $\sum_i \dcft_i,$ the 2d OPE coefficient in eq. \eqref{eq:c123n} is of the same form as in eq. \eqref{eq:scalar-AdS4} with the CFT$_3$ scaling dimensions $\dcft_i$ replaced by 2d scaling dimensions $\dccft_i$. 

 This result can be used to compute OPE blocks of 2d modes starting from the 3d OPE block \eqref{eq:scalar-block}. We show in Appendix \ref{app:shadow} that the shadow three-point function in eq. \eqref{eq:scalar-block} is given in terms of the 3-point function \eqref{eq:plane-3-pt} with $\Delta_3 \rightarrow 3 - \Delta_3$ by
 \begin{equation}
      \langle\Ocft{\dcft_1}(x_1) \Ocft{\dcft_2}(x_2) \Ocft{3-\dcft_3}^S(x_3) \rangle =  -2^{2\Delta_3 - \frac{9}{2}} \frac{\Gamma(\Delta_3 - \frac{3}{2})}{\Gamma(3 - \Delta_3)} \langle\Ocft{\dcft_1}(x_1) \Ocft{\dcft_2}(x_2) \Ocft{3-\dcft_3}(x_3) \rangle .
 \end{equation}
 Combined with the normalization of the OPE block in eq. \eqref{eq:shadow-norm}, we find
 \begin{equation}
 \label{eq:3d-block-int}
      \Ocft{\dcft_1}(t_1, \vec{z}_1)   \Ocft{\dcft_2}(t_2, \vec{z}_2) \supset N(\dcft_3) \int dt_3 \int d^2 \vec{z}_3 \langle   \Ocft{\dcft_1}(t_1, \vec{z}_1)   \Ocft{\dcft_2}(t_2, \vec{z}_2)   \Ocft{3 - \dcft_3}(t_3, \vec{z}_3) \rangle \Ocft{\dcft_3}(t_3, \vec{z}_3),
 \end{equation}
 where
 \begin{equation}
     N(\dcft_3) = -\frac{2^{2\dcft_3 - \frac{9}{2}}\Gamma(\dcft_3)}{\pi^3\Gamma(\frac{3}{2}-\dcft_3)}.
 \end{equation}

In order to reduce this to an OPE block of $\mathfrak{sl}(2,\mathbb{C})$ modes, we first express $\Ocft{\dcft_3}$ in terms of the analytically continued modes $\cftm{\dcft_3}{\lambda-1}$,
\begin{equation}
\label{eq:inverse}
    \Ocft{\dcft_3}(t_3, \vec{z}_3) = \frac{1}{2\pi i}\int_{-i\infty}^{i\infty} d\lambda (-1)^{\lambda}\frac{\pi}{\sin\pi(\lambda - 1)} (t_3+i\epsilon)^{\lambda - 1} \cftm{\dcft_3}{\lambda-1}(\vec{z}_3).
\end{equation}
It is straightforward to check that \eqref{eq:inverse} is equivalent to the definition \eqref{eq:Carroll-exp1} assuming that the contour can be closed in the right-hand complex $\lambda$ plane without picking up any contribution from the contour at infinity. 
Substituting \eqref{eq:inverse} into \eqref{eq:3d-block-int} and applying \eqref{eq:carr-cft-m} and \eqref{eq:ccft-carr} to relate the CFT$_3$ and celestial modes, we immediately find\footnote{We define $\Ocft{3-\dcft_3}^{(-\lambda)}$ by analytically continuing $\Ocft{3-\dcft_3}^{(n)}$, namely $\int_{-\infty}^{\infty} dt_3 t^{\lambda - 1} \Ocft{3 - \dcft_3}(t_3,\vec{z_3}) = 2\pi i \cftm{3-\dcft_3}{-\lambda}(\vec{z}_3)$.}
\begin{equation}
\label{eq:2d-block}
\begin{split}
      \Occft{\dcft_1-s_1-1}(\vec{z}_1)   \Occft{\dcft_2-s_2-1}(\vec{z}_2) 
      &\supset \frac{N(\dcft_3)}{2\pi^4} \frac{\cos\pi\dcft_3}{(3 - 2\dcft_3)}\int_{-i\infty}^{i\infty} d\lambda \int d^2 \vec{z}_3 (-1)^{\lambda} \\
      &\times \langle   \Occft{\dcft_1-s_1-1}(\vec{z}_1)   \Occft{\dcft_2-s_2-1}(\vec{z}_2)   \Occft{3 - \dcft_3-\lambda}( \vec{z}_3) \rangle \Occft{\dcft_3+\lambda-1}(\vec{z}_3).
      \end{split}
\end{equation}
In the case where $\dcft_3 \in \mathbb{Z}$, \eqref{eq:2d-block} is an $\mathfrak{sl}(2,\mathbb{C})$ block involving operator exchanges on the principal series. Deforming the contour in the complex $\lambda$ plane will allow us to pick out exchanges of dimensions determined by the poles of the integrand \cite{Dolan:2003hv,Dolan:2011dv}. 

The three-point function of 2d modes was already computed in \eqref{eq:2dccft}. Since the OPE coefficient in \eqref{eq:2dccft} is proportional to a delta function in dimensions, the integral over $\lambda$ automatically projects the RHS of \eqref{eq:2d-block} onto the \textit{residue} of $\Occft{\dccft_3}$ at 
\begin{equation}
\delta_3 = \delta_1 + \delta_2 - 2
\end{equation}  
to leading order in the Carrollian limit.
Putting everything together, we finally obtain the OPE of $\mathfrak{sl}(2,\mathbb{C})$ modes
\begin{equation}
\label{eq:2d-block-fin}
    \begin{split}
         \Occft{\delta_1}(\vec{z}_1)\Occft{\delta_2}(\vec{z}_2) &\supset \int d^2 \vec{z}_3 \frac{\mathcal{B}_{\dcft_1\dcft_2\dcft_3}(\dccft_1,\dccft_2)}{|z_{12}|^{2\delta_1 + 2\delta_2 - 4} |z_{13}|^{4 - 2\delta_2}|z_{23}|^{4 - 2\delta_1}} {\rm Res}_{\delta_3 =\delta_1 + \delta_2 - 2}\Occft{\delta_3}(\vec{z}_3).
    \end{split}
\end{equation}
Here, the normalization is inherited from the CFT$_3$ OPE block dual to tree-level scalar scattering in AdS$_4$. It is proportional to the coefficient in eq. \eqref{eq:c123n} with $\delta_3 \rightarrow2 - \delta_3$ and $\Delta_3 \rightarrow 3 - \Delta_3$, and takes the form\footnote{Note that in eq. \eqref{eq:c123n} we can replace $s_1 + s_2 + s_3 = -(\dccft_1 + \dccft_2 + \dccft_3) + \dcft_1 + \dcft_2 + \dcft_3 - 3$.}
\begin{equation}
\mathcal{B}_{\dcft_1\dcft_2\dcft_3}(\dccft_1,\dccft_2) = (-1)^{\dccft_1 + \dccft_2}  \mathcal{S}(\dcft_3)\mathcal{C}_{\dcft_1,\dcft_2,3-\dcft_3}(\delta_1,\delta_2,4-\dccft_1-\dccft_2), 
\end{equation}
with
\begin{equation}
    \mathcal{S}(\dcft_3) = (-1)^{\dcft_3}\frac{2^{2\Delta_3 - \frac{11}{2}} \cos\pi\Delta_3 \Gamma(\Delta_3)}{\pi^6 \Gamma(\frac{5}{2} - \Delta_3)} .
\end{equation}

Let us summarize our results so far. We started with the OPE block of two scalars in CFT$_3$, took the Carrollian limit and obtained a CFT$_2$ OPE block of a tower of operators with dimensions labeled by integers. The key step in this process was to write the CFT$_3$ three-point function in the representation \eqref{eq:3d-leadingc} and expand it in the Carrollian limit \textit{before} evaluating the integrals \eqref{eq:int-tr}. In taking the Carrollian limit, we have discarded any distributional terms. That such terms would appear can be seen by rescaling $\alpha_i \rightarrow \alpha_i/c$ in \eqref{eq:3d-leadingc} and applying the stationary phase approximation. The leading terms after this rescaling will then be distributional and will lead to a 2d OPE block where the standard CFT$_2$ three-point functions are replaced by celestial three-point functions. 

In \cite{deGioia:2024yne} we have shown that, after accounting for the normalization in \eqref{eq:carroll-spin}, the distributional terms (characteristic of the electric Carroll sector) dominate over the power law ones (characteristic of the magnetic Carroll sector). Here, we seem to find magnetic Carroll sector correlators at the same order in $c$ (recall that $\mathcal{C}$ does not depend on $c$). There is, however, no contradiction since here we treated the $c$-dependence as a distribution (see eq. \eqref{eq:c-distr}). Had we not, the Carrollian correlator associated with \eqref{eq:3d-leadingc} would multiply $c^{\sum_i \Delta_i -3}$ which vanishes as $c \rightarrow 0$ provided that $\sum_i \dcft_i > 3$. Interestingly, treating \eqref{eq:c-distr} as a distribution automatically projects the RHS of \eqref{eq:2d-block} precisely onto the exchange implied by collinear factorization of massless scalars in 4d flat space.  This will continue to be the case later when we will analyze the OPE blocks of currents and the stress tensor.  In contrast, in \cite{Guevara:2021abz} it had to be \textit{assumed} that the 2d OPE blocks contain the exchanges implied by collinear factorization. 

We also note that while the electric Carroll correlators vanish when operators are inserted on the same time slice \cite{deGioia:2024yne}, the magnetic sector ones do not. This is not surprising, since the magnetic sector correlators are not constrained by standard momentum conservation in the putative bulk dual, for a discussion of this point, see \cite{Bagchi:2025vri}. 

One surprising feature of the dimensional reduction is that the RHS of \eqref{eq:2d-block-fin} involves a residue of $\mathcal{O}_3$. At first one may think that this is precisely related to the normalization of 2d conformal primary wavefunctions which imply that conformally soft operators are obtained from residues of celestial amplitudes. In fact, we will see later on that the 2d OPE coefficients have additional poles at both positive and negative integer dimensions which one has to renormalize away in order to obtain a perfect match with celestial CFT. We will comment on this further at the end of Section \ref{sec:soft-gluon-thm}.

\section{Towers of 4d soft theorems from CFT$_3$ OPE blocks}
\label{sec:soft-tower}

The 3d CFT correlators involving a spin-$\ell$ operator and two scalars were worked out in \cite{Costa:2011mg} using the embedding space formalism.  It takes the form
\begin{equation}
\label{eq:lss}
    \langle O^{\ell}_{\Delta_1}(x_1) \Ocft{\Delta_2}(x_2) \Ocft{\Delta_3}(x_3) \rangle  = c_{\ell 12}\frac{\left[ (Z_1 \cdot P_2) (P_3 \cdot P_1) - (Z_1 \cdot P_3) (P_2 \cdot P_1) \right]^{\ell}}{(-P_1 \cdot P_2)^{\beta_{12} + \frac{\ell}{2}} (-P_2 \cdot P_3)^{\beta_{23} + \frac{\ell}{2}} (-P_3 \cdot P_1)^{\beta_{31} + \frac{\ell}{2}}}.
\end{equation}
Here $P_i= P(x_i)$ and $Z_i = Z(x_i)$ are respectively null momenta and polarization tensors in the embedding space $\mathbb{R}^{2,3}$ with metric ${\rm diag}(-1,-1,1,1,1)$. The form of \eqref{eq:lss} is essentially fixed by 3d conformal invariance which corresponds to Lorentz invariance and homogeneity under scaling in the embedding space. 
We review the basic ideas leading to this formula in Appendix \ref{sec:embedding}. 

It will be convenient to parameterize 
\begin{equation}
\label{eq:LPi}
\begin{split}
 P(t, z) = \sqrt{2}\left(\frac{1-t^2+z\bar z}{2},t,\frac{z+\bz}{2},\frac{-i(z-\bz)}{2},\frac{1+t^2-z\bar z}{2}\right)
 \end{split}
\end{equation}
and the polarization tensors $Z(t,z)$ as
\begin{equation}
\begin{split}
Z^{+}(t,z,\bz) = \p_z P(t,z,\bz) , \quad Z^{-}(t,z,\bz) = \p_{\bz} P(t,z,\bz), \quad Z^{t}(t,z,\bz) = -\p_t P(t,z,\bz).
    \end{split}
\end{equation}
In this case,
\begin{equation}
\label{eq:par}
\begin{split}
  P_i \cdot P_j &=  t_{ij}^2 - |z_{ij}|^2,\\
 Z_i^+ \cdot P_j &= -\bz_{ij}, \\
 Z_i^- \cdot P_j &= -z_{ij},
    \end{split}
\end{equation}
and the polarization tensors are orthonormal.
We will see that the correlators \eqref{eq:lss} are completely determined in terms of scalar three-point functions with shifted dimensions. Consequently, we will be able to directly apply the results of Section \ref{sec:dim-red-scalar} to compute the OPEs of the 2d modes of a 3d current-scalar and the stress tensor-scalar blocks. In this section, we will show that these reproduce exactly the towers of soft gluon and graviton theorems first derived in \cite{Hamada:2018vrw, Guevara:2021abz} with completely different methods.

\subsection{Tower of soft photon theorems}
\label{sec:soft-gluon-thm}

Let $J$ be a spin-$1$ operator. We will be interested in the case where $J$ is a conserved current, which in 3d further implies its dimension $\Delta_1 = 2$. We can use \eqref{eq:lss} to compute its OPE block with a scalar of dimension $\Delta$. Substituting the parameterization \eqref{eq:par} into \eqref{eq:lss} with $\ell = 1$ and $Z = Z^+$, we find
\begin{equation}
\label{eq:current-block}
\begin{split}
    J^+(x_1) \Ocft{\Delta}(x_2) &= c_{JO\widetilde{O}} \int d^3 x_3 \left[ - \frac{\bz_{12}}{(-t_{12}^2 + |z_{12}|^2)^{\beta_{12} + 1/2}(-t_{13}^2 + |z_{13}|^2)^{\beta_{13} - 1/2}(-t_{23}^2 + |z_{23}|^2)^{\beta_{23} + 1/2}} \right.\\
    &\left. + \frac{\bz_{13}}{(-t_{12}^2 + |z_{12}|^2)^{\beta_{12} - 1/2}(-t_{13}^2 + |z_{13}|^2)^{\beta_{13} + 1/2}(-t_{23}^2 + |z_{23}|^2)^{\beta_{23} + 1/2}} \right] \Ocft{\Delta}(x_3).
    \end{split}
\end{equation}
Up to the numerators which are time independent, we recognize in the brackets the \textit{scalar} three point function of operators with shifted dimensions $\Delta_2 = \Delta \rightarrow \Delta_2 + 1$ and $\Delta_3 = 3 - \Delta \rightarrow \Delta_3 + 1$ respectively and OPE coefficient stripped off. 
The (spin-$\ell$)-scalar-scalar OPE coefficient dual to the tree-level photon-scalar-scalar Witten diagram in AdS$_4$ was computed in \cite{Costa:2014spinning} (see eq. 131 therein) and is given by\footnote{Note that unlike in \cite{Costa:2014spinning}, our operators do not have unit normalization.}
\begin{equation}
\label{eq:spin-3-pt-coeff}
    c_{J\Ocft{\dcft_2} \Ocft{\dcft_3}} =  \frac{2^{\frac{\Delta_1 + \Delta_2 + \Delta_3}{2} - 3}}{2^{1-\ell}\pi^3}  g c^{-1} \frac{\Gamma(\beta_{12} + \frac{\ell}{2})\Gamma(\beta_{13} + \frac{\ell}{2})\Gamma(\beta_{23} + \frac{\ell}{2})\Gamma(\frac{\sum_i \Delta_i - 3 + \ell}{2})\Gamma(\Delta_1) (\ell + \dcft_1 - 1)}{(\Delta_1 - 1)\Gamma(\dcft_1 + \ell)\Gamma(\Delta_1 - \frac{1}{2})\Gamma(\Delta_2 - \frac{1}{2})\Gamma(\Delta_3 - \frac{1}{2})},
\end{equation}
where $\beta_{ij}$ was given in eq. \eqref{eq:alpha}. In this section we will use this formula with $\dcft_1 = 2$ and $\ell = 1.$
The OPE coefficient appearing in the OPE block \eqref{eq:current-block} is related to that in eq. \eqref{eq:spin-3-pt-coeff} by a  coefficient dependent on $\dcft$, namely
\begin{equation}
    c_{JO\widetilde{O}} = n_J(\dcft) c_{J\Ocft{\dcft}\Ocft{3-\dcft}}.
\end{equation}
Here $n(\dcft)$ is determined following the same approach in Appendix \ref{app:shadow}, but its precise form will be irrelevant here.

As a result, we can relate the current-scalar-scalar OPE block to the 3-scalar OPE block
\begin{equation}
\label{eq:current-weight-shift}
    \begin{split}
          J^+(x_1) \Ocft{\Delta}(x_2) &=  \int d^3 x_3 W_{JO}(\Delta_2, \Delta_3) \langle \Ocft{\Delta_1 = 2}(x_1) \Ocft{\Delta_2}(x_2) \Ocft{\Delta_3}(x_3) \rangle\left. \right|_{\Delta_2 = \Delta, \Delta_3 = 3 - \Delta}\Ocft{\Delta}(x_3), 
    \end{split}
\end{equation}
where $W_{JO}$ is the weight-shifting operator
\begin{equation}
\label{eq:weight-shift-def}
\begin{split}
     W_{JO}(\Delta_2, \Delta_3)&= c_{JO\widetilde{O}} \mathcal{D} \frac{1}{c_{2\Delta_2\Delta_3}}\\
    & = \sqrt{2} n_J(\dcft) \left(\left(\Delta_2 - \frac{1}{2}\right)\left (\beta_{13} - \frac{1}{2}\right) \bz_{12}e^{\p_{\Delta_2}} - \left(\Delta_3 - \frac{1}{2}\right)\left (\beta_{12} - \frac{1}{2}\right) \bz_{13} e^{\p_{\Delta_3}}\right)
     \end{split}
\end{equation}
and we defined
\begin{equation}
\label{eq:ws-bb}
    \mathcal{D} \equiv \bz_{12} e^{\p_{\Delta_2}} - \bz_{13} e^{\p_{\Delta_3}}.
\end{equation}
Note that the three-point function appearing in this formula now includes the OPE coefficient (see eq. \eqref{eq:plane-3-pt}). 
We prove eq. \eqref{eq:weight-shift-def} in Appendix \ref{eq:proof}. 
Upon accounting for the normalization relating CFT$_3$ and Carrollian/celestial operators, eq.  \eqref{eq:current-weight-shift} can be used to compute the OPE block of $\mathfrak{sl}(2,\mathbb{C})$ modes $\mathcal{J}^+_{1-s_1}$ and $\Occft{\dccft_2}$ of $J^+$ and $\Ocft{\Delta}$ by adapting the results of Section \ref{sec:dim-red-scalar}. We find
\begin{equation}
\label{eq:JO-2d-int}
\begin{split}
    &\mathcal{J}^+_{1 - s_1}(\vec{z}_1) \Occft{\dccft_2}(\vec{z}_2) = (-1)^{\dccft_1 + \dccft_2 - \dcft}\frac{i\cos\pi\dcft}{\sqrt{2}\pi^4(3 - 2\dcft)} n_J(\dcft)\int d\dccft_3 \int d^2 \vec{z}_3 \\
    & ~~~\left[ \left(\left(\beta_{13} - \frac{1}{2}\right) \bz_{12}e^{\p_{\dcft_2}} - \left(\beta_{12} - \frac{1}{2}\right) \bz_{13} e^{\p_{\dcft_3}}\right) \frac{\mathcal{C}_{2, \dcft_2, \dcft_3}(\dccft_1,\dccft_2,\dccft_3)}{|z_{12}|^{\delta_1 + \delta_2 - \delta_3} |z_{13}|^{\dccft_1+\dccft_2 - \dccft_3}|z_{23}|^{\dccft_2+\dccft_3-\dccft_1}} \right]_{\substack{\dcft_2 = \dcft, \dcft_3 = 3 - \dcft} }\\
    &\times \delta(-s_1+\delta_2+\dccft_3-2) (-s_1+\delta_2+\dccft_3-2) \Occft{2-\dccft_3}(\vec{z}_3).
    \end{split}
\end{equation}

We have therefore reduced the computation of the 2d OPE blocks associated with a spin-1 current and a scalar in CFT$_3$ to the action of a simple weight-shifting operator 
\begin{equation}
\label{eq:2d-wsop}
    \mathcal{W}_{JO}(\dcft_2, \dcft_3) = \left(\frac{\Delta_{32}}{2} + \frac{1}{2}\right) \bz_{12}e^{\p_{\dcft_2}} - \left(\frac{\Delta_{23}}{2} + \frac{1}{2}\right) \bz_{13} e^{\p_{\dcft_3}}, \quad \Delta_{23} = 2\Delta - 3
\end{equation}
on a 2d OPE block of scalars. To evaluate this, we first make the same change of variables 
\begin{equation}
\label{eq:ch-var}
    z_3 = z_2 + t z_{12}, \quad \bz_{3} = \bz_2 + \bar{t} \bz_{12}
\end{equation}
employed in \cite{Guevara:2021abz} in the computation of celestial (2d) OPE blocks of gluons and gravitons. The actions of $\bz_{12} e^{\p_{\Delta_2}}$ and $\bz_{13} e^{\p_{\Delta_3}}$ on the scalar three-point function in eq. \eqref{eq:JO-2d-int} can be verified by explicit computation to be equal up to a sign. As a result, keeping only the leading term in the $z_{12} \rightarrow 0$ expansion for fixed $\bz_{12}$ and evaluating the integral with\footnote{This is different from the choice of contour made in \cite{Guevara:2021abz}. One can show that other choices will lead to OPE coefficients that differ from those in eq. \eqref{eq:JOc} by ratios of trigonometric functions. For integer dimensions, these different choices will lead to the same number of poles.} $t \in [0,1]$ we obtain 
\begin{equation}
\label{eq:JO-fin}
    \begin{split}
        \mathcal{J}^+_{1 - s_1}(\vec{z}_1) \Occft{\dccft_2}(\vec{z}_2) &= \frac{1}{z_{12}}c^{2d}_{JO}(s_1, \dccft_2) \int d\bar{t} (1 - \bar{t})^{\dccft_2 - 1} \bar{t}^{- s_1 - 1} {\rm Res}_{\delta_3 =\dccft_2-s_1} \Occft{\dccft_3}(z_2, \bz_{2} + \bar{t} \bz_{12}) + \mathcal{O}(z_{12}^0),
    \end{split}
\end{equation}
with\footnote{Note that $\left(\beta_{13} + \beta_{12} - 1 \right) = \dcft_1 - 1 = 1$ since $\dcft_1 = 2.$}
\begin{equation}
\label{eq:JOc}
    c^{2d}_{JO}(s_1,\dccft_2) = (-1)^{1-s_1 + \dccft_2 - \dcft}\frac{ig n_J(\dcft) \cos\pi\dcft}{2\sqrt{2}(2\dcft-3)\sin \pi s_1 \sin \pi \dccft_2 } .
\end{equation}
The OPE coefficient exhibits some unexpected poles for $s_1, \delta_2 \in \mathbb{Z}$. These are \textit{not} the poles characteristic to celestial amplitudes of conformally soft modes \cite{Pate:2019mfs}. In order to obtain the expected OPE block of conformally soft modes, one has to further renormalize the operators, 
\begin{equation}
\label{eq:renormalization}
\begin{split}
   \mathcal{J}^+_{1 - s_1}(\vec{z}_1) &\rightarrow {\rm Res}_{\dccft_1 = 1- s_1}(\dccft_1 - 1 + s_1 )  \mathcal{J}^+_{\dccft_1}(\vec{z}_1),\\
    \Occft{\dccft_2}(\vec{z}_2) & \rightarrow {\rm Res}_{\dccft_2 = -s_2}(\dccft_2 + s_2 )  \Occft{\dccft_2}(\vec{z}_2).
   \end{split}
\end{equation}
The OPE blocks of these residues then precisely reproduce the tower of soft photon theorems, which can be seen by expanding eq. \eqref{eq:JO-fin} in powers of $\bz_{12}$ and evaluating the $\bar{t}$ integrals. In particular, one can compare eq. \eqref{eq:JO-fin} with the 2d celestial OPE blocks derived in \cite{Himwich:2021dau} finding perfect agreement -- cf. eq. 3.21 therein with $h_1 = \frac{2 - s_1}{2}, \bar{h}_1 = -\frac{s_1}{2}$, $h_2 = \bar{h}_2 = \frac{\Delta}{2}$ and $p = 0$ . 

We do not fully understand why the operators we obtain upon dimensional reduction do not exactly coincide with the operators appearing in celestial CFT. The same poles will continue to appear in all 2d OPE blocks that we will compute in this paper. In each case, they can be consistently eliminated by the same renormalization prescription in eq. \eqref{eq:renormalization}. We conclude this section by presenting some possible reasons for this apparent mismatch. 

Firstly, while we already accounted for the normalization expected to relate CFT$_3$ and celestial operators, we have assumed that operators in the magnetic Carroll sector are related to celestial operators by the same normalization as those in the electric Carroll sector. This may not be the case. Secondly, we restricted our analysis to an expansion of the CFT$_3$ around the same time slice. We have shown in \cite{deGioia:2024yne} that the expansions of the correlators around time slices of the Lorentzian cylinder differing by $\pi$ in global time (which correspond to in/out configurations) differ by phases. Taking linear combinations of such operators may, therefore, introduce trigonometric functions that cancel the poles we encountered. Finally, we would like to point out that \cite{Guevara:2021abz} started with a standard 2d CFT OPE block and also had to renormalize their operators to reproduce the collinear limit implied by 4d scattering amplitudes. From now on we will employ the renormalization prescription in eq. \eqref{eq:renormalization} and leave a better understanding of why it is necessary to future work.

\subsection{Tower of soft graviton theorems}

The 3d OPE block of the stress tensor with a scalar primary can be straightforwardly derived from \eqref{eq:lss} with $\ell = 2$. The OPE coefficient corresponding to a correlator dual to a tree-level AdS$_4$ Witten diagram of a graviton and two scalars is given by \eqref{eq:spin-3-pt-coeff} with $\ell = 2$ and $\Delta_1 = 3$, namely
\begin{equation}
    c_{T\Delta_2 \Delta_3} =  2^{\frac{\Delta_1 + \Delta_2 + \Delta_3}{2} - 2} \kappa c^{-1} \frac{\Gamma(\beta_{12} + 1)\Gamma(\beta_{13} + \frac{1}{2})\Gamma(\beta_{23} + 1)\Gamma(\frac{\sum_i \Delta_i - 1}{2})}{\pi^3 (\Delta_1 - 1)\dcft_1 \Gamma(\Delta_1 - \frac{1}{2})\Gamma(\Delta_2 - \frac{1}{2})\Gamma(\Delta_3 - \frac{1}{2})}, \quad \Delta_1 = 3.
\end{equation}
 We further focus on the contribution dual to a positive helicity graviton in AdS$_4$. This OPE block is related to the current OPE block \eqref{eq:current-weight-shift} by a double-copy relation,
   \begin{equation}
       W_{TO}(\Delta_2, \Delta_3) = c_{TO\widetilde{O}} \mathcal{D}^2 \frac{1}{c_{3\Delta_2 \Delta_3}},
   \end{equation}
   where $\mathcal{D}$ was defined in \eqref{eq:ws-bb}. 
   We prove in Appendix \ref{eq:proof2} that the associated component of the stress tensor-scalar block can be rewritten as
\begin{equation}
\label{eq:st-weight-shift}
 \begin{split}
          T^+(x_1) \Ocft{\Delta}(x_2) &= \int d^3 x_3 W_{TO}(\Delta_2, \Delta_3) \langle \Ocft{\Delta_1 = 3} \Ocft{\Delta_2} \Ocft{\Delta_3} \rangle\left. \right|_{\Delta_2 = \Delta, \Delta_3 = 3 - \Delta}\Ocft{\Delta}(x_3),
    \end{split}
    \end{equation}
    where 
    \begin{equation}
    \begin{split}
        W_{TO}&(\Delta_2,\Delta_3) = \frac{2 n_T(\dcft)}{\dcft_1(\dcft_1 - 1)}\left(\left(\Delta_2 - \frac{1}{2}\right)\left(\Delta_2 + \frac{1}{2}\right)\beta_{13}(\beta_{13} - 1)\bz_{12}^2 e^{2\p_{\Delta_2}}\right.\\
        &- \left. 2\left(\Delta_2 - \frac{1}{2}\right)\left(\Delta_3- \frac{1}{2}\right)\beta_{13}\beta_{12}\bz_{12}\bz_{13}e^{\p_{\Delta_2}}e^{\p_{\Delta_3}} 
        + \left(\Delta_3 - \frac{1}{2}\right)\left(\Delta_3 + \frac{1}{2}\right)\beta_{12}(\beta_{12} - 1)\bz_{13}^2 e^{2\p_{\Delta_3}} \right)
        \end{split}
    \end{equation}
    and $n_T(\dcft)$ is defined by
    \begin{equation}
        c_{TO\widetilde{O}} = n_T(\dcft) c_{TO_{\dcft} O_{3-\dcft}}.
    \end{equation}
   Here, we found this as a simple consequence of the  structure \eqref{eq:lss} determining the three-point function of a spin $\ell$ operator with two scalars, which is uniquely fixed by $\mathfrak{so}(2,3)$ symmetry. The same result can also be derived from the double-copy relating a graviton-scalar-scalar vertex to a photon/gluon-scalar-scalar vertex in the embedding space $\mathbb{R}^{3,2}$ or equivalently AdS$_4$. We present an alternative derivation of this position space double copy relation from a dual AdS$_4$ point of view in Appendix \ref{eq:double-copy}. Similar double-copy relations for \textit{momentum-space} CFT correlators in a flat space limit were discussed in \cite{Farrow:2018yni,Li:2022tby,Herderschee:2022ntr}.  

Once again, the OPE block of 2d $\mathfrak{sl}(2,\mathbb{C})$ modes of the 3d stress tensor with a 2d mode of the scalar can be directly deduced from the dimensional reduction of the scalar OPE block derived in Section \ref{sec:dim-red-scalar}. We find
\begin{equation}
\label{eq:TO-2d-int}
\begin{split}
    \mathcal{T}^+_{2 - s_1}(\vec{z}_1) \Occft{\dccft_2}(\vec{z}_2) &= (-1)^{\dccft_1 + \dccft_2 - \dcft}\frac{\cos\pi\dcft}{12\pi^4(3 - 2\dcft)} n_T(\dcft)  \\
    &\times \left[ \int d\dccft_3 \int d^2 \vec{z}_3 \mathcal{W}_{TO}(\dcft_2, \dcft_3) \frac{\mathcal{C}_{2, \dcft_2, \dcft_3}(\dccft_1,\dccft_2,\dccft_3)}{|z_{12}|^{\delta_1 + \delta_2 - \delta_3} |z_{13}|^{\dccft_1+\dccft_2 - \dccft_3}|z_{23}|^{\dccft_2+\dccft_3-\dccft_1}} \right]_{\dcft_2 = \dcft,\dcft_3 = 3 - \dcft} \\
    &\times \delta(-s_1+\delta_2+\dccft_3) (-s_1+\delta_2+\dccft_3) \Occft{2-\dccft_3}(\vec{z}_3),
    \end{split}
\end{equation}
where
\begin{equation}
    \mathcal{W}_{TO}(\dcft_2, \dcft_3) = \left(\Delta(\Delta - 1)\bz_{13}^2 e^{2\p_{\dcft_3}}- 2\Delta(3 - \Delta) \bz_{12}\bz_{13}e^{\p_{\dcft_2}}e^{\p_{\dcft_3}}+ (3 - \Delta)(2 - \Delta)\bz_{12}^2 e^{2\p_{\dcft_2}} \right).
\end{equation}
Making the same change of variables in \eqref{eq:ch-var} and acting with the corresponding weight-shifting operators, we obtain the simple result 
\begin{equation}
\label{eq:Tsc-2d}
    \begin{split}
       \mathcal{T}^+_{2 - s_1}(\vec{z}_1) \Occft{\dccft_2}(\vec{z}_2) &=  \frac{\bz_{12}}{z_{12}} c^{2d}_{TO}(s_1, \dccft_2) \int d\bar{t} (1 - \bar{t})^{\delta_2} \bar{t}^{- s_1} {\rm Res}_{\delta_3 = 2+ \dccft_2 -s_1} \Occft{\dccft_3}(z_2, \bz_{2} + \bar{t} \bz_{12}) + \mathcal{O}(z_{12}^0),
    \end{split}
\end{equation}
with 
\begin{equation}
    c^{2d}_{TO}(s_1,\dccft_2) = (-1)^{1-s_2 + \dccft_2 - \dcft}\frac{  \kappa n_T(\dcft) \cos\pi\dcft}{4 (2\dcft-3) \sin \pi s_1 \sin \pi \dccft_2 } .
\end{equation}
Upon redefining the operators according to \eqref{eq:renormalization} and expanding in $\bz_{12}$, we recover the tower of soft graviton theorems of \cite{Hamada:2018vrw}. We can also compare \eqref{eq:Tsc-2d} to the celestial OPE blocks derived in \cite{Himwich:2021dau}. Up to normalization, we see that this formula reproduces eq. 3.21 there, with $h_1 = \frac{4 - s_1}{2}, \bar{h}_1 = -\frac{s_1}{2}$, $h_2 = \bar{h}_2 = \frac{\Delta}{2}$ and $p = 1$.

\section{Celestial higher-spin symmetry algebras from CFT$_3$}
\label{eq:higher-spin}

In this section we will derive the OPE blocks of the 2d modes of CFT$_3$ currents $(\Delta = 2, \ell = 1)$ and the stress tensor $(\Delta = 3, \ell = 2)$ and their shadows with respectively $(\Delta = 1, \ell = 1)$ and $(\Delta = 0, \ell = 2)$. The contribution to these 3d OPE blocks from current and stress tensor exchanges are respectively determined by the three-point functions
\begin{equation}
\label{eq:current-3pt-main}
\langle J(P_1, Z_1) J(P_2, Z_2) \widetilde{J}(P_3, Z_3)\rangle = c_{JJ\widetilde{J}} \frac{2 V_1 V_2 V_3 + V_1 H_{23} + V_2 H_{13}}{(-P_1 \cdot P_2)^{3/2}(-P_1\cdot P_3)^{1/2}(-P_2\cdot P_3)^{1/2}}
\end{equation}
and
\begin{equation}
    \label{eq:stress-3pt}
   \langle T(P_1, Z_1) T(P_2, Z_2) \widetilde{T}(P_3, Z_3)\rangle =  c_{TT\widetilde{T}} \frac{ \sum_{i = 1}^{10} \alpha_i G_2^{(i)}(P_i; Z_i)}{(-P_{1}\cdot P_2)^{3}}.
\end{equation}
These three point functions were derived in \cite{Costa:2011mg} -- for a concise review see Appendix \ref{sec:embedding}. Here 
\begin{equation}
\label{eq:structures}
\begin{split}
  V_i &=  \frac{(Z_i \cdot P_j) (P_i \cdot P_k) - (Z_i \cdot P_k) (P_i \cdot P_j)}{\sqrt{(-P_i \cdot P_j)(-P_i \cdot P_k)(-P_j\cdot P_k)}} , \\
  H_{ij} &= Z_i \cdot Z_j - \frac{(Z_i \cdot P_j) (Z_j \cdot P_i)}{P_i \cdot P_j}, \quad i, j \in \{1,2,3\}
  \end{split}
\end{equation}
and $G_3^{(i)}$ are given in terms of the building blocks in \eqref{eq:stress-tensor-structures}. For abelian currents, the coefficient $c_{JJ\widetilde{J}}$ can be shown to vanish by symmetry. This closely parallels the vanishing of the photon three-point vertex in 4d flat space. We hence focus on non-abelian currents in which case $c_{JJ\widetilde{J}}$ will be proportional to antisymmetric structure constants of the associated Lie algebra. The number of independent coefficients $\alpha_i$ is reduced to 2 in CFT$_3$ by the stress tensor conservation constraints. Since both three-point functions receive contributions from different conformally covariant structures, it is not immediately clear which structures or what choices of coefficients correspond to the three point AdS-Witten diagram of gluons and gravitons in AdS$_4$. In this section, we will therefore express our results in terms of   $c_{JJ\tilde{J}}$ and $c_{TT\widetilde{T}}$.

We will be interested in the OPE block of positive helicity currents or the stress tensor. Setting $Z_1 = Z_1^+, Z_2 = Z_2^+$ and $Z_3=Z_3^-$ and using the parameterizations \eqref{eq:par}, we find
\begin{equation}
\begin{split}
    V_1^+ &= \frac{1}{(-t_{12}^2 + |z_{12}|^2)^{1/2}(-t_{13}^2 + |z_{13}|^2)^{1/2}(-t_{23}^2 + |z_{23}|^2)^{1/2}}\left[ -\bz_{12}(t_{13}^2 - |z_{13}|^2) + \bz_{13}(t_{12}^2 - |z_{12}|^2) \right],\\
      V_2^+ &= \frac{1}{(-t_{12}^2 + |z_{12}|^2)^{1/2}(-t_{13}^2 +|z_{13}|^2)^{1/2}(-t_{23}^2 + |z_{23}|^2)^{1/2}}\left[ \bz_{21}(t_{23}^2 - |z_{23}|^2) - \bz_{23}(t_{12}^2 - |z_{12}|^2) \right],\\
        V_3^- &= \frac{1}{(-t_{12}^2 + |z_{12}|^2)^{1/2}(-t_{13}^2 + |z_{13}|^2)^{1/2}(-t_{23}^2 + |z_{23}|^2)^{1/2}}\left[-z_{31}(t_{23}^2 - |z_{23}|^2) + z_{32}(t_{31}^2 - |z_{31}|^2) \right],
        \end{split}
\end{equation}
and
\begin{equation}
\label{eq:Hij}
\begin{split}
    H_{12}^{++} &= \frac{\bz_{12}^2}{t_{12}^2 - |z_{12}|^2},\\
    H_{13}^{+-} &= 1 + \frac{|z_{13}|^2}{t_{13}^2 - |z_{13}|^2} = \frac{t_{13}^2}{t_{13}^2 - |z_{13}|^2}, \\
    H_{23}^{+-} &= 1 + \frac{|z_{23}|^2}{t_{23}^2 - |z_{23}|^2} = \frac{t_{23}^2}{t_{13}^2 - |z_{13}|^2}.
    \end{split}
\end{equation}
We consider the projection of the three-point function on the negative helicity polarization tensor $Z_3^-$ because we are interested in the positive helicity exchanges in the positive helicity gluon and graviton OPE blocks.\footnote{To establish this result, we use completeness of polarization tensors. In general, the OPE blocks will also receive contributions from negative-helicity operators, as well as longitudinal modes.}

We have already encountered $V_1$ as the weight shifting operator in \eqref{eq:current-3pt}. In particular, when acting on scalar three-point functions (with OPE coefficient stripped off), $V_i$ become
\begin{equation}
\label{eq:V-weight-shift}
    V_1^+ = \bz_{12} e^{\p_{\Delta_2}} - \bz_{13} e^{\p_{\Delta_3}}, \quad  V_2^+ = \bz_{12} e^{\p_{\Delta_1}} + \bz_{23} e^{\p_{\Delta_3}}, \quad V_3^- =   -z_{13} e^{\p_{\Delta_1}} + z_{23} e^{\p_{\Delta_2}}.
\end{equation}
When acting on scalar three-point functions, $H_{ij}$ become the weight-shifting operators 
\begin{equation}
\label{eq:H-weight-shift}
\begin{split}
    H_{12}^{++} &= -\bz_{12}^2 e^{\p_{\Delta_1}} e^{\p_{\Delta_2}}, \\
     H_{13}^{+-} &= 1 - |z_{13}|^2 e^{\p_{\Delta_1}} e^{\p_{\Delta_3}},\\
      H_{23}^{+-} &= 1 - |z_{23}|^2 e^{\p_{\Delta_2}} e^{\p_{\Delta_3}}.
    \end{split}
\end{equation}

As before, thanks to \eqref{eq:V-weight-shift} and \eqref{eq:H-weight-shift}, we will be able to apply the decomposition of scalar 3d OPE blocks into 2d OPE blocks to perform a similar decomposition of the 3d OPE blocks of currents and stress tensors. It will be important to realize that, in the Carrollian limit, $H_{13}^{+-}$ and $H_{23}^{+-}$ become suppressed by a factor of $c^{2}$ with respect to $H_{12}^{++}$. This is evident from \eqref{eq:Hij}, and it is again crucial that the Carrollian limit is taken \textit{before} applying the integral transforms \eqref{eq:int-tr}.  
As a result, we find that terms in the OPE block multiplying $H_{13}^{+-}$ and $H_{23}^{+-}$ will give $\mathfrak{sl}(2,\mathbb{C})$ operators in with dimensions higher by two units for each such factor. These contributions should correspond to higher-dimension operators in the 4d flat space Lagrangian \cite{Pate:2019lpp}. We will see in the next section that $V_{2}^+ V_3^-$ will also contain terms suppressed by a factor of $c^{2}$ in the Carrollian limit. Our focus will be to recover the pure Yang-Mills and Einstein gravity celestial OPEs, so these terms will have to be identified and removed from our analysis. 

\subsection{$S$ algebra from CFT$_3$ current OPE block}

We would like to identify the leading term in the Carrollian limit of
$V_1^+ V_2^+ V_3^-$. We can do this by expressing this weight-shifting operator in terms of $H_{13}^{+-}$ and $H_{23}^{+-}$. We first write
\begin{equation}
\label{eq:v23j}
\begin{split}
    V_2^+ V_3^- &= \bz_{12} z_{31} e^{2\p_{\Delta_1}} - \bz_{12} z_{32} e^{\p_{\Delta_1}} e^{\p_{\Delta_2}} + \frac{\bz_{23}}{\bz_{13}} (H_{13}^{+-} - 1) - (H_{23}^{+-} - 1)\\
    &= -\frac{\bz_{12}}{\bz_{31}}(H_{13}^{+-} - 1)e^{\p_{\Delta_1}} e^{-\p_{\Delta_3}}  + \frac{\bz_{12}}{\bz_{32}}(H_{23}^{+-} - 1) e^{\p_{\Delta_1}} e^{-\p_{\Delta_3}}+ \frac{\bz_{23}}{\bz_{13}} (H_{13}^{+-} - 1) - (H_{23}^{+-} - 1)\\
    &= \frac{\bz_{12}^2}{\bz_{31}\bz_{32}} e^{\p_{\Delta_1}} e^{-\p_{\Delta_3}} + \frac{\bz_{12}}{\bz_{13}} + \mathcal{O}(H^{+-}_{13}, H^{+-}_{23}) = \frac{\bz_{12}}{\bz_{31}}\left(\frac{\bz_{12}}{\bz_{32}} e^{\p_{\Delta_1}} e^{-\p_{\Delta_3}} - 1 \right) + \mathcal{O}(H^{+-}_{13}, H^{+-}_{23}).
    \end{split}
\end{equation}
Consequently, the leading contribution to the current three-point function in the Carrollian limit can be expressed in terms of the scalar three-point function as 
\begin{equation}
\label{eq:JJ-weight-shift}
\begin{split}
    J^+(x_1) J^+(x_2) &= \int d^3 x_3 W_{JJ} \left[\langle \Ocft{2}(x_1) \Ocft{2}(x_2) \Ocft{\Delta_3}(x_3) \rangle \right] \left. \right|_{\Delta_3 = 1}J^+(x_3), \\
    W_{JJ}
    &=  c_{J^+J^+ \widetilde{J}^-}V_1^+ (V_{2}^+ V_3^-)\frac{1}{c_{\Delta_1 \Delta_2 \Delta_3}} \left. \right|_{\Delta_1 = \Delta_2 = 2, \Delta_3 = 1} \\
    &=  c_{J^+J^+ \widetilde{J}^-}\frac{c\pi^4}{2g} \bz_{21} e^{\p_{\Delta_3}}.
    \end{split}
\end{equation}

Note that a priori $V_1^+ (V_2^+ V_3^-)$ contains a linear combination of 4 weight-shifting operators, however, three of the coefficients turn out to vanish after setting $\dcft_1 = \dcft_2 = 2, \dcft_3=1$. We spell out the details of the short calculation leading to \eqref{eq:JJ-weight-shift} in Appendix \ref{eq:proof3}. The action of the weight-shifting operator $\frac{\bz_{12}}{\bz_{32}} e^{\p_{\Delta_1}} e^{-\p_{\Delta_3}}$ on the dimensionally reduced scalar three-point function can be shown by explicit computation to the equal to -1. As a result, the computation of \eqref{eq:JJ-weight-shift} reduces to multiplication by $\frac{\bz_{12}}{\bz_{31}}$ and acting with $V_1^+$ on the scalar three-point function. The latter is the differential operator already encountered in Section \ref{sec:soft-tower}. We saw there that, upon performing the dimensional reduction, $-\bz_{12} e^{\p_{\dcft_1}}$ and $\bz_{13} e^{\p_{\dcft_3}}$ have the same action on the 2d three-point function of scalars \eqref{eq:2dccft}. Following the same steps outlined in Section \ref{sec:soft-tower}, we then find
\begin{equation}
\label{eq:current-2d-block}
    \begin{split}
       \mathcal{J}_{1 - s_1}^+(\vec{z}_1) \mathcal{J}^+_{1 - s_2}(\vec{z}_2) &=  \frac{1}{z_{12}} c^{2d}_{JJ}(s_1, s_2) \int d\bar{t} (1 - \bar{t})^{-s_2 - 1} \bar{t}^{- s_1 - 1} {\rm Res}_{\dccft_3 = 1-s_1-s_2}\mathcal{J}^+_{\dccft_3}(z_2, \bz_{2} + \bar{t} \bz_{12}) + \mathcal{O}(z_{12}^0),
    \end{split}
\end{equation}
with 
\begin{equation}
    c^{2d}_{JJ}(s_1,s_2)  =  c_{J^+J^+\widetilde{J}^-} (-1)^{s_1+s_2+1}\frac{i\pi^4}{3 \sin \pi s_1 \sin \pi s_2 }. 
\end{equation}
After redefining the 2d primaries as explained in Section \ref{sec:soft-tower}, this result remarkably agrees with the celestial OPE block of gluons \cite{Fan:2019emx, Pate:2019lpp, Guevara:2021abz}. The derivation of the $S$ algebra then follows directly from \cite{Guevara:2021abz, Strominger:2021lvk}. 

We emphasize that this result follows from the general form of the current three-point function \eqref{eq:current-3pt-main} which applies to a generic CFT$_3$. In particular, we did not need to assume that the CFT admits a dual gravity description, although it was important to renormalize the operators according to \eqref{eq:carroll-spin} derived under the assumption that the CFT$_3$ three-point correlators correspond to contact three-point diagrams in AdS$_4$.

An alternative computation relies on identifying the contribution to the CFT$_3$ current three-point function dual to a $++-$ gluon three-point vertex in AdS. We review this derivation in Appendix \ref{eq:double-copy}. We would expect to recover eq. \eqref{eq:JJ-weight-shift} from the contribution to the current three-point function from the tree-level AdS-Witten diagrams of two positive and one negative helicity gluon. In terms of weight-shifting operators, the Yang-Mills (YM) three-point vertex in eq. \eqref{eq:flat-vertex} becomes
\begin{equation}
\label{eq:ws-YM-holo}
    V_{YM}^{++-} = \bz_{13} e^{\p_{\Delta_3}} - \bz_{12} e^{\p_{\Delta_1}}.
\end{equation}
While the differential operators appearing in eq. \eqref{eq:JJ-weight-shift} seem rather different, the two computations end up giving the same result. The reason for this is that the second bracket in eq. \eqref{eq:v23j} has a trivial action on the scalar three-point function.

\subsection{$w_{1+\infty}$ algebra from CFT$_3$ stress tensor OPE block}

We now turn to the stress tensor 3-point function given in eq. \eqref{eq:stress-3pt}. We argued before that structures involving $H_{13}^{+-}$ and $H_{23}^{+-}$ will be suppressed in the Carrollian limit. Consequently, only three of the structures in \eqref{eq:stress-tensor-structures} can contribute to the stress tensor 3-point function to leading order in this limit, namely
\begin{equation}
\label{eq:3-structures}
    (V_{1}^+)^2 (V_2^+)^2 (V_3^-)^2, \quad H_{12}^{++} V_{1}^+ V_2^+ (V_3^-)^2, ~~~ {\rm and} ~~~ (H_{12}^{++})^2( V_3^-)^2.
\end{equation}
The first structure is related to the square of the weight-shifting operator appearing in the YM case whose leading term in the Carrollian limit follows from a computation analogous to that in \eqref{eq:JJ-weight-shift}. We find
\begin{equation}
\begin{split}
    &(V_1^+)^2 (V_2^+)^2 (V_3^-)^2 \frac{1}{c_{\dcft_1\dcft_2\dcft_3}}\left. \right|_{\Delta_1 = \Delta_2 = 3, \Delta_3 = 0} = \frac{3c \pi^4}{8g} \bz_{12}^2 e^{2\p_{\Delta_3}}.
    \end{split}
\end{equation}
The other two structures involve products of the YM structure and $H_{12}^{++} V_3^-$ whose leading term in the Carrollian limit we evaluate now. 

First notice that
\begin{equation}
\begin{split}
    H_{12}^{++} V_3^- &= \bz_{12}^2 e^{\p_{\Delta_1}} e^{\p_{\Delta_2}} e^{-\p_{\Delta_3}} \left( -\frac{1}{\bz_{31}} |z_{13}|^2 e^{\p_{\Delta_1}} e^{\p_{\Delta_3}} + \frac{1}{\bz_{32}} |z_{23}|^2 e^{\p_{\Delta_2}}e^{\p_{\Delta_3}}\right)\\
    &=\frac{\bz_{12}^3}{\bz_{31} \bz_{32}} e^{\p_{\Delta_1}} e^{\p_{\Delta_2}} e^{-\p_{\Delta_3}}  + \mathcal{O}(H_{13}^{+-},H_{23}^{+-}).
    \end{split}
\end{equation}
It can be shown by explicit computation that the actions of the remaining two structures on the inverse scalar three-point coefficient vanishes,
\begin{equation}
    (H_{12}^{++}V_3^-)(V_1^+ V_2^+ V_3^-) \frac{1}{c_{\dcft_1\dcft_2\dcft_3}}\left. \right|_{\Delta_1 = \Delta_2 = 3, \Delta_3 = 0} = (H_{12}^{++}V_3^-)^2 \frac{1}{c_{\dcft_1\dcft_2\dcft_3}}\left. \right|_{\Delta_1 = \Delta_2 = 3, \Delta_3 = 0} = 0.
\end{equation}
Furthermore, these operators will continue to vanish when applied to properly normalized correlators. 
Consequently, only the first structure in \eqref{eq:3-structures}  gives a nontrivial contribution to the dimensionally reduced OPE block of CFT$_3$ stress tensor components. 
To leading order in the Carrollian limit we then find 
\begin{equation}
\label{eq:cft-weight-shift}
\begin{split}
    T^+(x_1) T^+(x_2) &= \int d^3 x_3 W_{TT} \left[\langle \Ocft{3} \Ocft{3} \Ocft{\Delta_3} \rangle \right] \left. \right|_{\Delta_3 = 0}T^+(x_3), \\
    W_{TT} &= c_{T^+T^+\widetilde{T}^-} \frac{3c \pi^4}{8g} \bz_{12}^2 e^{2\p_{\Delta_3}}.
    \end{split}
\end{equation}
This weight-shifting operator is manifestly proportional to the square of that in eq. \eqref{eq:JJ-weight-shift}.

To obtain the OPE blocks of sl$(2,\mathbb{C})$ modes of the stress tensor, we follow steps identical to those described in Section \ref{sec:soft-tower}, yielding\footnote{Note that to establish these relations one has to use the appropriate normalization relating \textit{spinning} CFT$_3$ primaries to Carollian/celestial ones, cf. eq. \eqref{eq:carroll-spin}.}
\begin{equation}
\label{eq:stress-2d-block}
    \begin{split}
        \mathcal{T}^+_{2 - s_1}(\vec{z}_1) \mathcal{T}^+_{2 - s_2}(\vec{z}_2) &=  \frac{\bz_{12}}{z_{12}} c^{2d}_{TT}(s_1, s_2) \int d\bar{t} (1 - \bar{t})^{-s_2} \bar{t}^{- s_1} {\rm Res}_{\dccft_3 = 4-s_1-s_2}\mathcal{T}^+_{\dccft_3}(z_2, \bz_{2} + \bar{t} \bz_{12}) + \mathcal{O}(z_{12}^0),
    \end{split}
\end{equation}
with 
\begin{equation}
    c^{2d}_{TT}(s_1,s_2) = c_{T^+T^+\widetilde{T}^-} (-1)^{s_1+s_2}\frac{\pi^4}{16 \sin \pi s_1 \sin \pi s_2 }.
\end{equation}
The OPE block \eqref{eq:stress-2d-block} remarkably agrees with the celestial OPE block of gravitons \cite{Guevara:2021abz}. The derivation of the celestial $w_{1 + \infty}$ algebra then follows immediately from \cite{Guevara:2021abz}. It is interesting that the ``mysterious'' $\dccft = 2$ conformally soft graviton naturally appears here as the first $(s = 0)$ mode in the tower of 3d CFT stress tensor modes.\footnote{Recall that the tower of soft theorems is associated with conformally soft gravitons of dimensions $\dccft \leq 1$, $\dccft \in \mathbb{Z}$.}

\section{Discussion}

 In this paper we presented further evidence that the flat space limit of AdS/CFT has the potential to capture all of the features of flat space holography. Our analysis focused on 3d CFTs where the flat space limit manifests itself as a Carrollian limit \cite{Duval:2014uva, PipolodeGioia:2022exe, Bagchi:2023fbj}. CFT$_3$ three-point functions of scalars, currents and the stress tensor are constrained by $\mathfrak{so}(3,2)$ symmetry to depend on a finite number of conformally covariant structures. In the Carrollian limit, the position-space three-point functions can be written as simple weight-shifting operators acting on the scalar correlators. These Carrollian correlators can then be decomposed into a tower of correlators of $\mathfrak{sl}(2,\mathbb{C})$ primaries of \textit{integer} dimensions. When applied to the 3d OPE blocks of a current and the stress tensor, this procedure allows one to obtain precisely the $\mathfrak{sl}(2,\mathbb{C})$ blocks of conformally soft gluons and gravitons used in \cite{Guevara:2021abz, Strominger:2021lvk} to compute their $S$ and $w_{1 + \infty}$ algebras. There are many directions of future research, a few of which we now outline. 

 Firstly, we have focused here on the leading contributions to the 3d OPE blocks in the flat space limit. It was shown in \cite{Taylor:2023ajd,Bittleston:2024rqe} that the $w_{1 + \infty}$ algebras of conformally soft gravitons admit a deformation proportional to the cosmological constant. Since the cosmological constant $\Lambda \sim R^{-2} \sim c^2$, we expect to be able to recover this deformation by analyzing the first subleading corrections to the Carrollian limit of OPE blocks involving the CFT$_3$ stress tensor. We can in fact already see from acting on \eqref{eq:3d-leadingc} with the appropriate weight-shifting operators that these corrections will produce an operator exchange of the correct dimension multiplied by an overleading pole in $\sim z_{12}^{-2}$. It remains to check that the associated OPE coefficient agrees with that computed in \cite{Taylor:2023ajd,Bittleston:2024rqe} using the Jacobi identity. We see, however, no reason why an associative CFT$_3$ (as CFTs with holographic AdS duals are usually assumed to be) should result in a theory obeying the standard axioms of CFT$_2$ and in particular the Jacobi identity, after dimensional reduction. 
 
 Relatedly, it will be important to understand from general principles which CFT properties should survive after taking the Carrollian limit and projecting to an $\mathfrak{sl}(2,\mathbb{C})$ basis.
Here we have assumed that the Carrollian limit of the 3d OPE blocks is taken \textit{first}, and in particular, prior to taking the 2d OPE limit. It would be interesting to analyze the interplay between the different orders of limits, with the goal of shedding light on the associativity of the celestial OPE. It will also be important to better understand the role of shadow transforms. Specifically, in \cite{deGioia:2023cbd} the leading and subleading conformally soft theorems appeared in a flat space limit of the shadow stress tensor Ward identities. Here we extracted the whole tower of conformally soft theorems from OPE blocks of CFT$_3$ currents and stress tensors. 

 Finally, our series of works \cite{PipolodeGioia:2022exe, deGioia:2023cbd, deGioia:2024yne} focused mostly on understanding how the rich symmetries of celestial conformal field theories arise in a limit of CFT$_3$. The underlying geometric picture and the source of these symmetry enhancements from the perspective of a bulk AdS$_4$ dual remain largely unclear. We hope to answer some of these questions in future work.

\section*{Acknowledgements}
We would like to thank Laurent Freidel, Daniele Pranzetti, Biswajit Sahoo, Kostas Skenderis, Andrew Strominger and Diandian Wang for discussions. We are grateful to Laurent Freidel, Monica Pate and Daniele Pranzetti for comments on a draft. A.R. was in part supported by the Simons Foundation through the Emmy Noether Fellows Program at Perimeter Institute (1034867, Dittrich). A.R. is grateful for the hospitality of Perimeter Institute where part of this work was carried out. L. G. acknowledges the support provided by FAPESP Foundation through the grant 2023/04415-2. Research at Perimeter Institute is supported in part by the Government of Canada through the Department of Innovation, Science and Economic Development and by the Province of Ontario through the Ministry of Colleges and Universities.

\appendix

\section{From $\mathfrak{so}(3,2)$ to 4d Poincar\'e}
\label{app:3d-algebra}

Consider a CFT$_3$ on the Lorentzian plane with metric given by
\be 
\label{eq:Euclidean-metric}
ds^2 = -dt^2 + dz d\bar{z},
\ee
we find that \eqref{eq:cft-gens} become 
\begin{equation}
\label{eq:3d-so32}
  \begin{split}
[T_{\mu}, \Ocft{\dcft}(x)] &= -i \left(\p_t, \p_z, \p_{\bz}\right) \Ocft{\dcft}(x), \\
[K_{t}, \Ocft{\dcft}(x)] &= - i \left(  2t \Delta + 2 t \bz\p_{\bz}+2t z\p_{z} + (t^2 + z\bz) \p_t\right) \Ocft{\dcft}(x),\\
[K_{z}, \Ocft{\dcft}(x)] &= -i \left(-t^2 \p_z - \bz^2 \p_{\bz} - t \bz \p_t - \Delta \bz\right) \Ocft{\dcft}(x), \\
[K_{\bz}, \Ocft{\dcft}(x)] &= -i\left(-t^2 \p_{\bz} - z^2 \p_{z} - t z \p_t - \Delta z\right)\Ocft{\dcft}(x)\\
[D,\Ocft{\dcft}(x)] &= -i\left(\Delta + z\p_z + \bz \p_{\bz} + t \p_t \right) \Ocft{\dcft}(x),\\
[J_{tz}, \Ocft{\dcft}(x)] &= -i \left(- t\p_z - \frac{1}{2} \bz \p_t \right)\Ocft{\dcft}(x), \quad [J_{t\bz}, \Ocft{\dcft}(x)] = -i \left( -t\p_{\bz} -\frac{1}{2} z \p_t \right)\Ocft{\dcft}(x),\\
[J_{z\bz}, \Ocft{\dcft}(x)] &= - \frac{i}{2}\left( \bz\p_{\bz} - z \p_{z} \right)\Ocft{\dcft}(x).
\end{split}
\end{equation}
It is straightforward to see that in the limit $c \rightarrow 0$ with $t = c u$ and $u$ as well as $c T_t, c K_t, c J_{tz}, c J_{t\bz}$ fixed, the $\mathfrak{so}(3,2)$ generators become 4d Poincar\'e generators.

\section{Derivation of eq. \eqref{eq:master-f}}
\label{app:derivation}

Let $\vec{x} \in \mathbb{R}^n$. In this section we prove the identity
\begin{equation}
\label{eq:id0}
\begin{split}
\frac{\Gamma(\beta_{12}) \Gamma(\beta_{13}) \Gamma(\beta_{23})}{|\vec{x}_{12}|^{2\beta_{12}} |\vec{x}_{13}|^{2\beta_{13}} |\vec{x}_{23}|^{2 \beta_{23}}} &= \prod_{i = 1}^3 \left( \int_0^{\infty} d\omega_i \right) \omega_1^{\beta_{23} - 1} \omega_{2}^{\beta_{13} - 1} \omega_3^{\beta_{12} - 1} e^{-\omega_1 |\vec{x}_{23}|^2-\omega_2 |\vec{x}_{13}|^2-\omega_3 |\vec{x}_{12}|^2 } \\
&= \prod_{i = 1}^3 \left( \int_0^{\infty} d\alpha_i  \alpha_i^{\Delta_i - 1} \right) \alpha^{-\frac{\Delta_1 + \Delta_2 + \Delta_3}{2}} \left(\frac{\alpha}{\pi} \right)^{n/2} \int d^n \vec{x} e^{-\sum_{i = 1}^3 \alpha_i |\vec{x} - \vec{x}_i|^2 }
\end{split}
\end{equation}
with $\beta_{ij}$ defined in \eqref{eq:alpha}. 
In the second line we have changed variables
\begin{equation}
\omega_1 = \frac{\alpha_2 \alpha_3}{\alpha}, \quad \omega_2 = \frac{\alpha_1 \alpha_3}{\alpha}, \quad \omega_3 = \frac{\alpha_2 \alpha_1}{\alpha}, \quad  \alpha = \sum_{i = 1}^3 \alpha_i,
\end{equation}
whose Jacobian is
\begin{equation}
\left| \frac{\p \omega}{ \p\alpha}\right| = \frac{\alpha_1\alpha_2 \alpha_3}{\alpha^3}
\end{equation}
and we used the identity
\begin{equation}
\label{eq:integral-id}
\begin{split}
\int d^n \vec{x} e^{-\sum_{i = 1}^3 \alpha_i |\vec{x} - \vec{x}_i|^2} &= \int d^n \vec{x} e^{-\alpha |\vec{x} - \sum_{i = 1}^3 \frac{\alpha_i}{\alpha} \vec{x}_i|^2 -\left(\frac{\alpha_1 \alpha_2}{\alpha} |\vec{x}_1 - \vec{x}_2|^2 + \frac{\alpha_1 \alpha_3}{\alpha} |\vec{x}_1 - \vec{x}_3|^2 + \frac{\alpha_2 \alpha_3}{\alpha} |\vec{x}_2 - \vec{x}_3|^2 \right) }\implies\\
&e^{-\left(\frac{\alpha_1 \alpha_2}{\alpha} |\vec{x}_1 - \vec{x}_2|^2 + \frac{\alpha_1 \alpha_3}{\alpha} |\vec{x}_1 - \vec{x}_3|^2 + \frac{\alpha_2 \alpha_3}{\alpha} |\vec{x}_2 - \vec{x}_3|^2 \right) } = \left(\frac{\alpha}{\pi} \right)^{n/2 }\int d^n \vec{x} e^{-\sum_{i = 1}^3 \alpha_i |\vec{x} - \vec{x}_i|^2}.
\end{split}
\end{equation}
The latter is derived by rewriting 
\begin{equation}
\begin{split}
\sum_{i = 1}^3 \alpha_i |\vec{x} - \vec{x}_i|^2  &= \alpha \left|\vec{x} - \sum_{i = 1}^3 \frac{\alpha_i}{\alpha} \vec{x}_i\right|^2 + \frac{\alpha_1 \alpha_2}{\alpha} |\vec{x}_1 - \vec{x}_2|^2 + \frac{\alpha_1 \alpha_3}{\alpha} |\vec{x}_1 - \vec{x}_3|^2 + \frac{\alpha_2 \alpha_3}{\alpha} |\vec{x}_2 - \vec{x}_3|^2. \\
 \end{split}
\end{equation}
This formula can be applied to Lorentzian CFT$_3$ 3-point functions by replacing $\omega_i \rightarrow \pm i\omega_i$ depending on the $i\epsilon$ prescription that defines the analytic continuation of the correlator past its lightcone singularities.

\section{Shadow three-point function}
\label{app:shadow}

In this appendix we compute the shadow transform of the scalar CFT$_3$ 3-point function \eqref{eq:plane-3-pt}. This is given by
\begin{equation}
\begin{split}
\langle\Ocft{\dcft_1}(x_1) \Ocft{\dcft_2}(x_2) \Ocft{3-\dcft_3}^S(x') \rangle &= \int d^3P_3 \frac{1}{(-P'\cdot P_3)^{3 - \Delta_3}}\langle\Ocft{\dcft_1}(P_1) \Ocft{\dcft_2}(P_2) \Ocft{\dcft_3}(P_3) \rangle\\
&=    \int dt_3 d^2 z_3 \frac{1}{(-(t' - t_3)^2 + |z' - z_3|^2)^{(3 - \Delta_3)}} \langle\Ocft{\dcft_1}(x_1) \Ocft{\dcft_2}(x_2) \Ocft{\dcft_3}(x_3) \rangle.
    \end{split}
\end{equation}

We again use the Mellin representation of the three-point function together with the identity \eqref{eq:integral-id} to write this as 
\begin{equation}
\begin{split}
  &\langle\Ocft{\dcft_1}(x_1) \Ocft{\dcft_2}(x_2) \Ocft{3-\dcft_3}^S(x') \rangle = c_{123} \frac{i^{-\frac{\Delta_1 + \Delta_2 + \Delta_3 + 1}{2}}}{\Gamma(\beta_{12})\Gamma(\beta_{13})\Gamma(\beta_{23})}  \prod_{i = 1}^3 \left( \int_0^{\infty} d\alpha_i  \alpha_i^{\Delta_i - 1} \right) \alpha^{-\frac{\Delta_1 + \Delta_2 + \Delta_3}{2}} \left(\frac{\alpha}{\pi} \right)^{3/2}  \\
&\times \frac{i^{-3 + \dcft_3} }{\Gamma(3 - \dcft_3)}\int dt_3 d^2 z_3 \int_0^{\infty} d\gamma_3 \gamma_3^{2 - \Delta_3} e^{i\gamma_3(-(t' - t_3)^2 + |z' - z_3|^2 + i\epsilon)} \int dt \int d^2z  e^{i\sum_{i = 1}^3 \alpha_i (-(t - t_i)^2 + |z - z_i|^2 + i\epsilon)}.
\end{split}
\end{equation}
We can now use the generalization of the identity \eqref{eq:integral-id} on two points to evaluate the $t_3$ and $z_3$ integrals, 
\begin{equation}
    \begin{split}
  &\langle\Ocft{\dcft_1}(x_1) \Ocft{\dcft_2}(x_2) \Ocft{3-\dcft_3}^S(x') \rangle =  \frac{c_{123}}{\Gamma(3 - \dcft_3)}\frac{ i^{-\frac{\Delta_1 + \Delta_2 - \Delta_3 + 1}{2}}}{\Gamma(\beta_{12})\Gamma(\beta_{13})\Gamma(\beta_{23})}  \prod_{i = 1}^3 \left( \int_0^{\infty} d\alpha_i  \alpha_i^{\Delta_i - 1} \right) \alpha^{-\frac{\Delta_1 + \Delta_2 + \Delta_3}{2}} \left(\frac{\alpha}{\pi} \right)^{3/2}  \\
&\times  i^{-5/2}\int_0^{\infty} d\gamma_3 \gamma_3^{2 - \Delta_3} \left(\frac{\pi}{(\gamma_3 + \alpha_3)} \right)^{3/2} \int dt \int d^2z  e^{i\sum_{i = 1}^2 \alpha_i (-(t - t_i)^2 + |z - z_i|^2 + i\epsilon) + i \frac{\alpha_3 \gamma_3}{\alpha_3 + \gamma_3}(-(t' - t)^2 + |z' - z|^2 + i\epsilon)}.
\end{split}
\end{equation}
We now make the change of variables 
\begin{equation}
    \gamma_3' = \frac{\alpha_3 \gamma_3}{\alpha_3 + \gamma_3} \implies \gamma_3 = \frac{\alpha_3 \gamma_3'}{\alpha_3 - \gamma_3'}, \quad d\gamma_3' = \frac{\alpha_3^2}{(\alpha_3 + \gamma_3)^2} d\gamma_3
\end{equation}
and evaluate the integral over $\alpha_3$ after changing the order of the $\alpha_3$ and $\gamma_3'$ integrals and shifting $\alpha_3 \rightarrow \alpha_3 + \gamma'_3$. We find
\begin{equation}
\begin{split}
    \langle\Ocft{\dcft_1}(x_1) \Ocft{\dcft_2}(x_2) \Ocft{3-\dcft_3}^S(x') \rangle &=  \frac{c_{123}}{\Gamma(3 - \dcft_3)}\frac{ i^{-\frac{\Delta_1 + \Delta_2 - \Delta_3 + 1}{2}}}{\Gamma(\beta_{12})\Gamma(\beta_{13})\Gamma(\beta_{23})}  \prod_{i = 1}^3 \left( \int_0^{\infty} d\alpha_i  \alpha_i^{\Delta_i - 1} \right) \alpha^{-\frac{\Delta_1 + \Delta_2 + \Delta_3}{2}}   \\
&\times  i^{-5/2}\left(\frac{\alpha}{\pi} \right)^{3/2} \int_0^{\alpha_3} d\gamma_3' \frac{\alpha_3^2}{(\alpha_3 - \gamma_3')^2} \left(\frac{\alpha_3 \gamma_3'}{\alpha_3 - \gamma_3'} \right)^{2 - \Delta_3} \left(\frac{\pi( \alpha_3 - \gamma_3')}{\alpha_3^2} \right)^{3/2}\\
&\times \int dt \int d^2z  e^{i\sum_{i = 1}^2 \alpha_i (-(t - t_i)^2 + |z - z_i|^2 + i\epsilon) + i \frac{\alpha_3 \gamma_3}{\alpha_3 + \gamma_3}(-(t' - t)^2 + |z' - z|^2 + i\epsilon)}\\
& =  \frac{c_{123}}{\Gamma(3 - \dcft_3)}\frac{ i^{-\frac{\Delta_1 + \Delta_2 - \Delta_3}{2}-3}}{\Gamma(\beta_{12})\Gamma(\beta_{13})\Gamma(\beta_{23})} \frac{\Gamma(\Delta_3 - \frac{3}{2})\Gamma(\beta_{12})}{\Gamma(\frac{\Delta_1 + \Delta_2  + \Delta_3 - 3}{2})} \\
&\times  \int_0^{\infty} d\alpha_1  \alpha_1^{\Delta_1 - 1} \int_0^{\infty} d\alpha_2 \alpha_2^{\Delta_2 - 1}\int_0^{\infty} \alpha_3^{3 - \Delta_3 - 1} \alpha^{-\frac{\Delta_1 + \Delta_2 - \Delta_3}{2}} \left(\frac{1}{\pi} \right)^{3/2}  \\
&\times \int dt \int d^2z  e^{i\sum_{i = 1}^2 \alpha_i (-(t - t_i)^2 + |z - z_i|^2 + i\epsilon) + i \alpha_3(-(t' - t)^2 + |z' - z|^2 + i\epsilon)}.\\
\end{split}
\end{equation}
We have relabeled the $\gamma_3'$ integration variable to $\alpha_3$. In the last line we can apply \eqref{eq:id0} with $\Delta_3 \rightarrow \widetilde{\dcft}_3 = 3 - \Delta_3$, hence we arrive at the final result
\begin{equation}
\begin{split}
     \langle\Ocft{\dcft_1}(x_1) \Ocft{\dcft_2}(x_2) \Ocft{3-\dcft_3}^S(x_3) \rangle &= -\frac{c_{123}}{\Gamma(3 - \dcft_3)}\frac{1}{\Gamma(\beta_{13})\Gamma(\beta_{23})} \frac{\Gamma(\Delta_3 - \frac{3}{2})}{\Gamma(\frac{\Delta_1 + \Delta_2  + \Delta_3 - 3}{2})} \Gamma(\widetilde{\beta}_{12})\Gamma(\widetilde{\beta}_{13})\Gamma(\widetilde{\beta}_{23})\\
     &\times \langle \Ocft{\dcft_1}(x_1) \Ocft{\dcft_2}(x_2) \Ocft{3-\dcft_3}(x_3) \rangle_{u.n.}\\
     &= -2^{\Delta_3 - \frac{3}{2}} \frac{\Gamma(\Delta_3 - \frac{3}{2})}{\Gamma(3 - \Delta_3)} \langle\Ocft{\dcft_1}(x_1) \Ocft{\dcft_2}(x_2) \Ocft{3-\dcft_3}(x_3) \rangle .
     \end{split}
\end{equation}
Here we defined 
\begin{equation}
    \widetilde{\beta}_{ij}(\Delta_1, \Delta_2, \Delta_3) = \beta_{ij}(\Delta_1, \Delta_2, 3 - \Delta_3)
\end{equation}
and the subscript u.n. denotes the correlator with the three-point coefficient stripped off.

\section{Review of CFT in the embedding space}
\label{sec:embedding}
 
The action of conformal symmetries in CFT$_d$ lifts to the linear action of the Lorentz group on a lightcone in the embedding space Mink$^{d,2}$. This allows one to  build conformal invariants in $d$-dimensions from Lorentz invariants in the $(d+2)$-dimensional embedding space. As shown in \cite{Costa:2011mg}, this equivalence can be successfully exploited to construct CFT correlators involving spinning operators, as we now review.

Consider CFT$_d$ primary operators with quantum numbers $(\Delta, \ell)$, where $\Delta$ is the conformal dimension and $\ell$ is the spin. These can be lifted to Lorentz tensors in Mink$^{d,2}$ which we denote by $\Phi_{A_1\cdots A_\ell}(P)$ obeying the following properties:  
 \begin{itemize}
 \item $P^2 = 0$, ie. these correspond to massless fields in the embedding space
 \item are homogeneous of degree $-\Delta$, ie. $\Phi_{A_1\cdots A_\ell}(\lambda P) = \lambda^{-\Delta} \Phi_{A_1\cdots A_\ell}(P)$
 \item are symmetric and traceless
 \item $P^{A_1} \Phi_{A_1 A_2 \cdots A_\ell} = 0$
 \end{itemize}
 It will be convenient to trade $\Phi_{A_1 \cdots A_{\ell}}$ for Lorentz scalars by contraction with embedding space polarization vectors $Z^A$. These obey the usual identities
 \be
 \label{eq:pol-prop}
 Z^2 = 0 = P(x) \cdot Z(x). 
 \ee
 The spin in then encoded through the homogeneity of the resulting polynomial $\Phi(P; Z)$ in $Z$, namely
 \be 
 \Phi(P; \lambda Z) = \lambda^{\ell} \Phi(P; Z).
 \ee
In the remainder of this appendix we review the construction of conformal two- and three-point functions of primary operators in terms of Lorentz invariants in the embedding space. 
 
 \subsection{Two-point functions of $(\Delta, \ell)$}
 \label{sec:scalars}
 
 We write 
 \be 
 \langle O_1^{(\Delta,\ell)} O_2^{(\Delta,\ell)}  \rangle = \frac{G_2(P_i; Z_i)}{(P_1 \cdot P_2)^{\Delta + \ell}}.
 \ee
 Gauge invariance, namely invariance under $Z \rightarrow Z + \alpha P$ together with $P\cdot \Phi = 0$ tells us that the building blocks for $O_i$ are 
 \be 
 C_{i AB} = Z_{iA} P_{iB} - Z_{iB} P_{iA}.
 \ee
 By \eqref{eq:pol-prop}, $C_i \cdot C_i = 0$, so the only invariant appearing in the two-point function must involve $C_1 \cdot C_{2}$, 
 \be
 \frac{C_1 \cdot C_2}{P_1 \cdot P_2} = \frac{2 (P_1 \cdot P_2) (Z_1\cdot Z_2) - 2 (Z_1 \cdot P_2) (Z_2 \cdot P_1)}{P_1 \cdot P_2} \equiv 2 H_{12}.
 \ee
 Homogeneity in $\Delta$ and $\ell$ then fixes the form of the two point function
 \be 
  \langle O_1^{(\Delta,\ell)}O_2^{(\Delta,\ell)}  \rangle \propto \frac{H_{12}^{\ell}}{(P_1 \cdot P_2)^{\Delta}}
 \ee
 up to normalization.
 
 \subsection{Three-point functions of spins $(\ell_1\ell_2\ell_3) = (00\ell)$}
 \label{sec:spin}
 
 We write
 \be
 \langle O_1 O_2 O_3^{\ell} \rangle \propto \frac{G_3(P_i; Z_i)}{(P_{1}\cdot P_2)^{\alpha_{12}} (P_{1}\cdot P_3)^{\alpha_{13}} (P_{2}\cdot P_3)^{\alpha_{23}} }, \quad \alpha_{ij} = \frac{1}{2}\left(\Delta_i + \Delta_j  - \Delta_k\right).
 \ee
 Using the results from the previous section, $O_3$ must be built from $C_{3AB}$ and since $P_3 \cdot C_3 = Z_3 \cdot C_3  = 0$, the only three-point Lorentz invariant that we can write down is
 \be 
 V_{3,12} = \frac{P_1 \cdot C_3 \cdot P_2}{\sqrt{(P_1 \cdot P_2)(P_1 \cdot P_3)(P_2\cdot P_3)}} = \frac{(Z_3 \cdot P_1) (P_2 \cdot P_3) - (Z_3 \cdot P_2) (P_1 \cdot P_3)}{\sqrt{(P_1 \cdot P_2)(P_1 \cdot P_3)(P_2\cdot P_3)}} \equiv V_3.
 \ee
 Homogeneity then fixes the form of the three-point function to be proportional to
 \be
 \langle O_1 O_2 O_3^{\ell} \rangle \propto \frac{(V_{3,12})^{\ell}}{(P_{1}\cdot P_2)^{\alpha_{12}} (P_{1}\cdot P_3)^{\alpha_{13}} (P_{2}\cdot P_3)^{\alpha_{23}} }.
 \ee
 
 \subsection{Three-point functions of arbitrary spin}
 
 We can now use the results of section \ref{sec:scalars} and \ref{sec:spin} to construct the three-point function of primary operators of arbitrary spin. We first note that we can either contract different $C_i$ with each other, or contract $C_i$ with $P_k, Z_k$ with $k \neq i$. One could for example consider
 \be 
 \begin{split}
 Z_1 \cdot C_3 \cdot Z_2 &= (Z_3 \cdot Z_1) (Z_2 \cdot P_3) - (Z_3\cdot Z_2) (Z_1 \cdot P_3),
 \end{split}
 \ee
 however, we see that this structure is not invariant under $Z_{1, 2} \rightarrow Z_{1,2} + P_{1,2}$ and is therefore not allowed. $Z_1 \cdot C_3 \cdot P_2$ has the same problem and so we are left with either contractions of $C_i$ with $P_k$ or contractions among different $C_i$. It is shown in \cite{Costa:2011mg} that these don't give rise to any new structures with respect to those encountered so far. Therefore, we can write 
 \be 
 \langle O_1^{\ell_1} O_2^{\ell_2} O_3^{\ell_3} \rangle \propto \frac{\prod_i V_i^{m_i} \prod_{i < j} H_{ij}^{n_{ij}}}{(P_{1}\cdot P_2)^{\alpha_{12}} (P_{1}\cdot P_3)^{\alpha_{13}} (P_{2}\cdot P_3)^{\alpha_{23}} },
 \ee
 where in the last line we used the symmetry properties 
 \be 
 H_{ij} = H_{ji}, \quad V_{i, jk} = -V_{i, kj}.
 \ee
 Homogeneity in $P_i$ and $Z_i$ then constrains the powers $m_i$, $n_{ij}$ for each $i = 1, 2, 3$
 \be 
 \ell_i = m_i + \sum_{j \neq i} n_{ij}.
 \ee
 Provided there are not further constraints from symmetry (ie. all spins are distinct) or conservation laws, we can count the number of independent 3-point structures $(\ell_1 \ell_2 \ell_3)$. This amounts to find $n_{ij} \leq 0$ that satisfy
 \be 
 \begin{split}
 n_{12} + n_{13} \leq \ell_1, \quad n_{12} + n_{23} \leq \ell_{2}, \quad n_{23} + n_{13} \leq \ell_3
 \end{split}
 \ee
 which amounts to counting the number of integer points inside a polyhedron defined by the equations above. An explicit formula is given in \cite{Costa:2011mg}. In this paper, we will only be interested in the stress tensor 3-point function. The stress tensor Ward identities impose further constraints which reduce the number of independent structures that can appear from 5 to 3 for $d > 3$ and from 4 to 2 in the case $d = 3$. This case is sufficiently simple that the distinct structures can be enumerated, which we do next. 

 \subsection{Current 3-point function}

 In the case where one or more operators are conserved, the number of conformal structures contributing to the three-point function is reduced due to the conservation law(s). In this appendix we review the structures contributing to a three-point function of conserved $\ell = 1$ currents \cite{Zhiboedov:2012bm,Costa:2011mg}. The three-point function of two conserved currents and one primary of dimension $\Delta$ and spin $\ell$ takes the form
 \begin{equation}
 \label{eq:current-3pt}
     \langle J_1(x_1) J_2(x_2) \mathcal{O}_{\Delta, \ell}(x_3) \rangle = \frac{\mathcal{T}_{\ell}(P_i, Z_i)}{(P_1 \cdot P_2)^{2 - \Delta/2} (P_1 \cdot P_3)^{\Delta/2}(P_2 \cdot P_3)^{\Delta/2}}
 \end{equation}
 with 
 \begin{equation}
     \mathcal{T}_1 = \lambda_1 V_1 V_2 V_3 + \lambda_2 V_1 H_{23} + \lambda_3 V_2 H_{13} + \lambda_4 V_3 H_{12}
 \end{equation}
 for even parity operators. 
 We will further consider the case where $\mathcal{O}_{\Delta,\ell}$ is a spin-1 conserved current (i.e. $\Delta = 2, \ell = 1$) or its shadow (i.e. $\Delta = 1, \ell = 1$). Note that the shadow is in general not a conserved operator, but this will not be relevant since we will only consider the constraints coming from the conservation of operators $1$ and $2$. As a result, one obtains relations among the $\lambda_i$. We can further consider the cases where the three-point function is symmetric or antisymmetric under the exchange of $1$ and $2$. In the former case one finds that $\lambda_i = 0$, while in the latter case one finds \cite{He:2023ewx} 
 \begin{equation}
 \label{eq:lambda-rel}
     \lambda_1 = (\Delta + 3) \lambda_2, \quad \lambda_3 = \lambda_3, \quad \lambda_4 = (\Delta - 1)\lambda_2.
 \end{equation}
For non-abelian currents transforming in the adjoint representation of a Lie algebra, the three-point function will be proportional to the structure constant which is antisymmetric. We can therefore have a non-vanishing current three-point function of the form \eqref{eq:current-3pt} with $\lambda_i$ related by \eqref{eq:lambda-rel}.

 \subsection{Stress tensor 3-point function}
 
 $(\ell_1, \ell_2, \ell_3) = (2,2,\ell)$. Imposing symmetry under exchange of $1$ and $2$ reduces the number of allowed structures to 10 (see appendix A of \cite{Costa:2011mg}):
 \be 
 \label{eq:stress-tensor-structures}
 \begin{split}
& \underbrace{ V_1^2 V_2^2 V_3^{\ell}}_{G_{\ell}^{(1)}}, \quad  (H_{13} V_2^2 V_1 + H_{23} V_1^2 V_2) V_3^{\ell - 1}, \quad H_{12} V_1 V_2 V_3^{\ell}, \quad (H_{13} V_2 + H_{23} V_1) H_{12} V_2^{\ell - 1}, \\
& H_{13} H_{23} V_1 V_2 V_3^{\ell - 2}, \quad H_{12}^2 V_3^{\ell}, \quad (H_{13}^2 V_2^2 + H_{23}^2 V_1^2) V_{3}^{\ell - 2}  \\
 & H_{12} H_{23} H_{13} V_3^{\ell - 2}, \quad (H_{13} H_{23}^2 V_1 + H_{23} H_{13}^2 V_2) V_3^{\ell - 3}, \quad \underbrace{ H_{13}^2 H_{23}^2 V_3^{\ell - 4}}_{G^{(10)}_{\ell}}.
 \end{split}
 \ee
 For $d > 3,$ these are a-priori all independent, and we can write
 \be 
 \label{eq:stress-three-pt-d}
 \langle T_1 T_2 \mathcal{O}_{\Delta,\ell} \rangle = \frac{ \sum_{i = 1}^{10} \alpha_i G_{\ell}^{(i)}(P_i; Z_i)}{(P_{1}\cdot P_2)^{\frac{d}{2}} (P_{1}\cdot P_3)^{\frac{d}{2}} (P_{2}\cdot P_3)^{\frac{d}{2}}}.
 \ee
 Further imposing the conservation of $1$ and $2$ yields 7 linearly independent relations among the coefficients $\alpha_i$, thereby reducing the number of independent coefficients to 3.  These relations can be found in \cite{Costa:2011mg}. We will be interested in the case where $\mathcal{O}_{\Delta, \ell}$ are either the stress tensor or its shadow, ie. $\ell = 2$  and $\Delta = 3$ and $\Delta = 0$ respectively. In the case of the stress tensor 3-point function in CFT$_3$, the conservation equation further reduces the number of independent linear combinations of structures to 2. These independent structures correspond to the stress tensors of free bosons and free fermion theories.

\section{Relating spinning and scalar OPE blocks}

\subsection{Proof of eq. \eqref{eq:current-weight-shift}}
\label{eq:proof}

We define the weight-shifting operators 
\begin{equation}
\label{eq:wieght-shift-bb}
    \mathcal{D}^{(2)} = \bz_{12} e^{\p_{\Delta_2}}, \quad \mathcal{D}^{(3)} = -\bz_{13} e^{\p_{\Delta_3}}.
\end{equation}
Then the three-point function appearing in the OPE block \eqref{eq:current-block} can be written as
\begin{equation}
\begin{split}
   & c_{J\Delta_2 \Delta_3} \left(\mathcal{D}^{(2)} + \mathcal{D}^{(3)} \right) \frac{1}{c_{\dcft_1\dcft_2\dcft_3}} \langle \Ocft{\Delta_1}(x_1) \Ocft{\Delta_2}(x_2) \Ocft{\Delta_3}(x_3) \rangle \\
    &= c_{J\Delta_2 \Delta_3}\left(\frac{1}{c_{\dcft_1\dcft_2+1\dcft_3}} \mathcal{D}^{(2)} + \frac{1}{c_{\dcft_1\dcft_2\dcft_3+1}}  \mathcal{D}^{(3)} \right) \langle \Ocft{\Delta_1}(x_1) \Ocft{\Delta_2}(x_2) \Ocft{\Delta_3}(x_3) \rangle.
    \end{split}
\end{equation}
It is straightforward to check using the definitions that
\begin{equation}
    \frac{c_{J\Delta_2\Delta_3}}{c_{\dcft_1\dcft_2+1\dcft_3}} = \frac{\sqrt{2}}{(\Delta_1 - 1)} (\Delta_2 - \frac{1}{2}) (\beta_{13} - \frac{1}{2}), \quad  \frac{c_{J\Delta_2\Delta_3}}{c_{\dcft_1\dcft_2\dcft_3+1}} = \frac{\sqrt{2}}{(\Delta_1 - 1)} (\Delta_3 - \frac{1}{2}) (\beta_{12} - \frac{1}{2})
\end{equation}
from which \eqref{eq:current-weight-shift} follows upon setting $\Delta_1 = 2$. 

In preparation for the dimensional reduction, we would like to write the 3d OPE block in terms of a weight-shifting operator acting on the appropriately normalized three-point function. We also find
\begin{equation}
   \left( \prod_{i = 1}^3 N_i^{\ell_i}\right) c_{J\Delta_2 \Delta_3} \left(\mathcal{D}^{(2)} + \mathcal{D}^{(3)} \right) \frac{1}{c_{\dcft_1\dcft_2\dcft_3}\left( \prod_{i = 1}^3 N_i^0\right) } = -\frac{i}{\sqrt{2}}\left((\beta_{13} - \frac{1}{2}) \mathcal{D}^{(2)} + (\beta_{12} - \frac{1}{2})\mathcal{D}^{(3)}\right),
\end{equation}
where $N_i^{\ell}$ is the net normalization relating CFT$_3$ and celestial modes given by the product of the relative prefactors in \eqref{eq:carroll-spin} and \eqref{eq:norm}.

\subsection{Proof of eq. \eqref{eq:st-weight-shift}}
\label{eq:proof2}

Using the definitions \eqref{eq:wieght-shift-bb}, we note that the graviton-scalar-scalar OPE block involves the three-point function
\begin{equation}
\label{eq:weight-shift-st}
\begin{split}
    &c_{T\Delta_2 \Delta_3}\left(\mathcal{D}^{(2)} + \mathcal{D}^{(3)} \right)^2 \frac{1}{c_{\Delta_1 \Delta_2 \Delta_3}} \langle \Ocft{\Delta_1}(x_1) \Ocft{\Delta_2}(x_2) \Ocft{\Delta_3}(x_3) \rangle\\
    &=  c_{T\Delta_2 \Delta_3}(\mathcal{D}^{(2)} + \mathcal{D}^{(3)})\left( \frac{1}{c_{\dcft_1\dcft_2+1\dcft_3}} \mathcal{D}^{(2)} + \frac{1}{c_{\dcft_1\dcft_2\dcft_3+1}}  \mathcal{D}^{(3)}\right)\langle \Ocft{\Delta_1}(x_1) \Ocft{\Delta_2}(x_2) \Ocft{\Delta_3}(x_3) \rangle\\
    &= c_{T\Delta_2 \Delta_3}\left( \frac{1}{c_{\dcft_1\dcft_2+2\dcft_3}} (\mathcal{D}^{(2)})^2 + \frac{2}{c_{\dcft_1\dcft_2+1\dcft_3+1}}  \mathcal{D}^{(3)} \mathcal{D}^{(2)} + \frac{1}{c_{\dcft_1\dcft_2\dcft_3 +2}}( \mathcal{D}^{(3)})^2\right)\\
    &\times \langle \Ocft{\Delta_1}(x_1) \Ocft{\Delta_2}(x_2) \Ocft{\Delta_3}(x_3) \rangle.
    \end{split}
\end{equation}
One can check the following identities hold
\begin{equation}
    \begin{split}
        \frac{c_{T\Delta_2\Delta_3}}{c_{\Delta_1 \Delta_2 + 2\Delta_3}} &= \frac{2}{\dcft_1(\Delta_1 - 1)} \left(\Delta_2 - \frac{1}{2} \right) \left(\Delta_2 + \frac{1}{2} \right) \beta_{13}(\beta_{13} - 1), \\
         \frac{c_{T\Delta_2\Delta_3}}{c_{\Delta_1 \Delta_2\Delta_3 + 2}} &= \frac{2}{\dcft_1(\Delta_1 - 1)} \left(\Delta_3 - \frac{1}{2} \right)\left(\Delta_3 + \frac{1}{2} \right) \beta_{12} (\beta_{12} -1), \\
         \frac{c_{T\Delta_2\Delta_3}}{c_{\Delta_1 \Delta_2 + 1\Delta_3+1}} &= \frac{2}{\dcft_1(\Delta_1 - 1)}  \left(\Delta_2 - \frac{1}{2} \right) \left(\Delta_3 - \frac{1}{2} \right) \beta_{13} \beta_{12} ,~~~ 
    \end{split}
\end{equation}
from which the result follows upon setting $\Delta_1 = 3$.

\subsection{Proof of eq. \eqref{eq:JJ-weight-shift}}
\label{eq:proof3}

We first use the identity
\begin{equation}
    \left(\frac{\bz_{12}}{\bz_{32}} e^{\p_{\Delta_1}} e^{-\p_{\Delta_3}} - 1 \right) \frac{1}{c_{\dcft_1\dcft_2\dcft_3}} = \frac{1}{c_{\dcft_1\dcft_2\dcft_3}} \left(\frac{\Delta_1 - \frac{1}{2}}{\Delta_3 - \frac{3}{2}} \frac{\beta_{23}-1}{\beta_{12}} \frac{\bz_{12}}{\bz_{32}} e^{\p_{\Delta_1}}e^{-\p_{\Delta_3}}  - 1\right).
\end{equation}
Composing this with $V_1^+ \frac{\bz_{12}}{\bz_{31}}$, we find
\begin{equation}
\label{eq:v123}
\begin{split}
 V_1^+ V_2^+ V_3^-  & \frac{1}{c_{\dcft_1\dcft_2\dcft_3}} = 2^{\frac{3 - \sum_i\dcft_i}{2}} \frac{c\pi^3 \bz_{12}}{g \bz_{23}} \frac{\Gamma(\dcft_1 - \frac{1}{2})\Gamma(\dcft_2 - \frac{1}{2})\Gamma(\dcft_3 - \frac{3}{2})}{\Gamma(\beta_{12} + \frac{1}{2})\Gamma(\beta_{13} - \frac{1}{2})\Gamma(\beta_{23} + \frac{1}{2})\Gamma(\frac{1}{2}(\dcft_1 + \dcft_2 + \dcft_3 -2))}\\
 &\times \left(\frac{2\dcft_3 - 3}{2\beta_{13} - 1}\left( -\bz_{23}(2\beta_{12} - 1)(2\dcft_3 - 1) e^{\p_{\dcft_3}}  + \bz_{12}(2\dcft_1 - 1)(1 - 2\beta_{23}) e^{\p_{\dcft_1}} \right) \right.\\
 &+ \left.\frac{\bz_{12}(2\dcft_2 - 1)}{\bz_{13}(1+2\beta_{12})} \left(\bz_{23} (2\beta_{12}+1)(2\dcft_3 -3)e^{\p_{\dcft_2}} + \bz_{21}(2\dcft_1 - 1)(1 - 2\beta_{23}) e^{\p_{\dcft_1}} e^{\p_{\dcft_2}} e^{-\p_{\dcft_{3}}} \right)\right).
 \end{split}
\end{equation}
Now note that for $\dcft_1 = \dcft_2 = 2, \dcft_3 = 1$, we have
\begin{equation}
    \beta_{12} = \frac{3}{2}, \quad \beta_{13} = \frac{1}{2}, \quad \beta_{23} = \frac{1}{2}.
\end{equation}
We hence see that the overall coefficient has a zero due to $\Gamma(\beta_{13} - \frac{1}{2})$ in the denominator. The only term that has a canceling pole is the first one in the second line. The result \eqref{eq:JJ-weight-shift} then follows immediately.

\section{CFT$_3$ correlators from AdS-Witten diagrams}
\label{eq:double-copy}

In holographic CFTs, the leading contributions at large $N$ to the current and stress tensor three-point functions are given by the \textit{tree-level} three-point gluon and graviton Witten diagrams in Yang-Mills theory and Einstein gravity in AdS$_4$ \cite{Witten:1998qj,Gubser:1998bc}. Uplifted to the embedding space, these are given by 
\begin{equation}
\label{eq:YM-amp}
    A_{\rm YM}^{\pm \pm \pm} = \Pi_{A}^{~~A_1 A_2 A_3} Z^+_{A_1} Z^+_{A_2} Z^-_{A_3} \int_{AdS_4} D^4X \left[\p_{X_A} K_{\Delta_1}(P_1;X) K_{\Delta_2}(P_2;X) K_{\Delta_3}(P_3;X) + {\rm perms.} \right]
\end{equation}
and
\begin{equation}
\label{eq:3-amplit}
\begin{split}
A_{\rm EG}^{\pm \pm \pm} &= \Pi_{AB}^{~~~A_1B_1;A_2B_2;A_3B_3} Z_{A_1B_1}^{\pm}(P_1) Z_{A_2 B_2}^{\pm}(P_2) Z^{\pm}_{A_3B_3}(P_3)\\
&\times \int_{AdS_4} D^4X \left[\p_{X_A} K_{\Delta_1}(P_1;X) \p_{X_B} K_{\Delta_2}(P_2;X) K_{\Delta_3}(P_3;X) + {\rm perms.} \right].
\end{split}
\end{equation}
Here $\Pi_{AB}^{~~~A_1A_2;B_1B_2;C_1C_2}$ is a projector that implements the appropriate index contractions appearing in the three-point graviton vertex. $K_{\Delta}(P_i; X)$ are scalar bulk-to-boundary propagators in AdS$_4$, which in terms of embedding space vectors $P, X$ take the form
\begin{equation}
\label{eq:btb}
K_{\Delta}(P; X) = \frac{C_{\Delta}}{(-P\cdot X)^{\Delta}}, \quad C_\Delta = \dfrac{2^{\Delta}\Gamma(\Delta)}{2\pi^{3/2}\Gamma(\Delta-\frac{1}{2})}.
\end{equation}
Consequently, \eqref{eq:YM-amp} and \eqref{eq:3-amplit} straightforwardly simplify to 
\begin{equation}
   A_{\rm YM} = V_{\rm YM}^{\pm\pm\pm} A_{\rm s}, \quad   A_{\rm EG}^{\pm \pm \pm} = V_{\rm EG}^{\pm\pm\pm} A_{\rm s},
\end{equation}
with
\begin{equation}
\label{eq:eg-vertex}
\begin{split}
 V_{\rm YM}^{++-} &= \Pi_{A}^{~~~A_1 A_2 A_3 } Z_{A_1}^{\pm}(P_1) Z_{A_2}^{\pm}(P_2) Z^{\pm}_{A_3}(P_3) \mathcal{P}_1^A  + {\rm perms.},\\
    V_{\rm EG}^{\pm \pm \pm}( \mathcal{P}_i,Z_i) &= \Pi_{AB}^{~~~A_1B_1;A_2B_2;A_3B_3} Z_{A_1B_1}^{\pm}(P_1) Z_{A_2 B_2}^{\pm}(P_2) Z^{\pm}_{A_3B_3}(P_3) \mathcal{P}_1^A \mathcal{P}_2^B + {\rm perms.}, \\
    \mathcal{P}_i^A &\equiv \frac{(\Delta_i - \frac{1}{2})}{2} P_i^A e^{\partial_{{\Delta}_i}}
    \end{split}
\end{equation}
the YM and Einstein gravity vertices and 
\begin{equation}
\label{eq:scalar-AdSW}
    A_s = \int_{AdS_4} D^4X K_{\Delta_1}(P_1;X) K_{\Delta_2}(P_2;X) K_{\Delta_3}(P_3;X)
\end{equation}
the scalar AdS Witten diagram. The gauge theory and gravity 3-point vertices naively take a complicated form, however, in flat spacetimes they can be simplified using momentum conservation and are furthermore related by a double-copy relation \cite{Bern:2010ue}. The double copy for AdS amplitudes and their CFT duals was discussed in \cite{Farrow:2018yni, Armstrong:2020woi,Zhou:2021gnu, Herderschee:2022ntr}. 

Naively, that AdS amplitudes also obey a double copy relation should follow from \eqref{eq:eg-vertex} which are gauge theory and gravity vertices in flat space (here the embedding space $\mathbb{R}^{3,2}$). However this is too fast, because the  ``momenta'' in \eqref{eq:eg-vertex} are in fact weight-shifting operators and a priori not conserved. 
On the other hand, integration by parts in \eqref{eq:3-amplit} shows that the weight-shifting operators are in fact related, up to boundary terms, by substitutions of the form
\begin{equation}
\label{eq:m-c}
    \mathcal{P}_1 = -\mathcal{P}_2 -\mathcal{P}_3.
\end{equation}
Hence, the vertex \eqref{eq:eg-vertex} can  only differ by boundary terms in AdS$_4$. By definitions, we also have 
\begin{equation}
\label{eq:orthpz}
Z_i^{\pm} \cdot \mathcal{P}_i = 0.
\end{equation}
Under the assumption that the boundary terms vanish, the vertex can then be further simplified using \eqref{eq:m-c} and \eqref{eq:orthpz}. 

In conclusion, \eqref{eq:eg-vertex} takes the same form as the momentum-space three-point graviton vertex in flat space subject to the replacement of the usual momenta with $\mathcal{P}_i$, which formally obey \eqref{eq:m-c} when acting on the scalar AdS-Witten diagram, and upon identifying $Z_i$ with the corresponding polarization tensors. As such, we expect it to be expressed via the double copy formula, in terms of the Yang-Mills three-point vertex \cite{Farrow:2018yni} 
\begin{equation}
\label{eq:flat-vertex}
  V_{\rm EG}(p_i;\epsilon_i) = (V_{\rm YM})^2,\quad  V_{\rm YM}(p_i,\epsilon_i) = (\epsilon_1\cdot\epsilon_2)(\epsilon_3 \cdot p_2) + {\rm cyclic}.
\end{equation}
subject to the replacements $\epsilon_i \rightarrow Z(P_i), \quad p_i \rightarrow \mathcal{P}_i$. Related observations have appeared in \cite{Raju:2012zr, Farrow:2018yni, Li:2022tby}. In these references, the double copy arises in a flat space limit of momentum space CFT correlators and the copy formula for CFT correlators involves replacements of $\epsilon_i, p_i$ by differential operators of a more complicated form. It would be interesting to clarify the relation between these different perspectives.  

We conclude that the holographic current three-point function in  CFT$_3$ can be related to a scalar primary correlator of dimensions $(\Delta_1, \Delta_2, \Delta_3) = (2,2,1)$ by
\begin{equation}
    \label{eq:current-c}
\begin{split}
\langle J^+(P_1) J^+(P_2) \widetilde{J}^-(P_3)\rangle &\propto V_{\rm YM}^{++-} (\mathcal{P}_i,Z_i) \langle O_{2}(P_1) O_{2}(P_2) O_{1}(P_3) \rangle .
\end{split}
\end{equation}
Similarly, the holographic stress tensor three-point function in CFT$_3$ is given in terms of the scalar correlator of primaries of dimensions $(\Delta_1, \Delta_2, \Delta_3) = (3,3,0)$ by
\begin{equation}
    \label{eq:stress-c}
\begin{split}
\langle T^+(P_1) T^+(P_2) \widetilde{T}^-(P_3)\rangle &\propto V^{++-}_{\rm EG}(\mathcal{P}_i,Z_i) \langle O_{3}(P_1) O_{3}(P_2) O_{0}(P_3) \rangle .
\end{split}
\end{equation}
Note that the weight-shifting operator in \eqref{eq:eg-vertex} depends on the normalization of the bulk-to-boundary propagators in \eqref{eq:btb}. It is therefore important that the three-point scalar correlator inherits the normalization from \eqref{eq:scalar-AdSW}, namely
\begin{equation}
    \label{eq:three-point-scalar}
    \langle O_{\Delta_1}(x_1) O_{\Delta_2}(x_2) O_{\Delta_3}(x_3)\rangle = \frac{c_{123}}{|x_1 - x_2|^{\Delta_1 + \Delta_2 - \Delta_3}|x_2 - x_3|^{\Delta_2 + \Delta_3 - \Delta_1}|x_1 - x_3|^{\Delta_1 + \Delta_3 - \Delta_2}}
\end{equation}
with
\begin{equation}
\label{eq:c123}
c_{123} = \frac{2^{\frac{\Delta_1 + \Delta_2 + \Delta_3}{2} - 3}\Gamma(\alpha_{12})\Gamma(\alpha_{13})\Gamma(\alpha_{23})\Gamma\left(\frac{\Delta_1 + \Delta_2 + \Delta_3 - 3}{2} \right)R}{2\pi^3 \Gamma(\Delta_1 - \frac{1}{2})\Gamma(\Delta_2  - \frac{1}{2})\Gamma(\Delta_3  - \frac{1}{2})},
\end{equation}
and $R$ is the AdS radius. In the main text we use a dual perspective, where we obtain identical formulas by considering a Carrollian limit (with $R$ identified with $c^{-1}$) of the CFT$_3$ position space correlators.

\bibliographystyle{utphys}
\bibliography{references}

\end{document}